\documentclass[aps,prd,showpacs,floatfix,nofootinbib,twocolumn]{revtex4-1}

\usepackage[version=3]{mhchem} 
\usepackage{enumerate}
	
\usepackage[bottom]{footmisc}
\usepackage{mathrsfs}
\usepackage{dcolumn}
\usepackage{bm}
\usepackage{amsmath}
\usepackage{amsthm}
\usepackage{amssymb}
\usepackage{graphicx}
\usepackage{footmisc}
\usepackage{nicefrac}
\usepackage{appendix}

\usepackage[normalem]{ulem}
\usepackage[capitalise]{cleveref}
\usepackage{cancel}
\usepackage[dvipsnames,table,xcdraw]{xcolor}

\newcommand{\eref}[1]{(\ref{#1})}
\newcommand{\eps }{\epsilon }
\usepackage{epstopdf}

\newcommand{\ben}{\begin{equation}}
\newcommand{\een}{\end{equation}}
\newcommand{\bea}{\begin{eqnarray}}
\newcommand{\eea}{\end{eqnarray}}

\def\ext{_{\rm ext}}

\def\en{_{\rm en}}

\def\bR{{\bf R}}

\begin{document}

\title{On the Numerical Solution of the Exact Factorization Equations}
\author{Graeme H. Gossel}
\affiliation{Department of Physics and Astronomy, Hunter College and the City University of New York, 695 Park Avenue, New York, New York 10065, USA}
\author{Lionel Lacombe}
\affiliation{Department of Physics and Astronomy, Hunter College and the City University of New York, 695 Park Avenue, New York, New York 10065, USA}
\author{Neepa T. Maitra}
\affiliation{Department of Physics and Astronomy, Hunter College and the City University of New York, 695 Park Avenue, New York, New York 10065, USA} 
\affiliation{The Physics Program and the Chemistry Program of the Graduate Center, City University of New York, 365 Fifth Avenue, New York, USA}

\date{\today}
\begin{abstract}
The exact factorization (EF) approach to coupled electron-ion dynamics recasts the time-dependent molecular Schr\"odinger equation as two coupled equations, one for the nuclear wavefunction and one for the conditional electronic wavefunction. The potentials appearing in these equations have provided insight into non-adiabatic processes, and new practical non-adiabatic dynamics methods have been formulated starting from these equations. Here we provide a first demonstration of a self-consistent solution of the exact equations, with a preliminary analysis of their stability and convergence properties. The equations have an unprecedented mathematical form, involving a non-Hermitian Hamiltonian, and so the usual numerical methods for time-dependent Schr\"odinger fail when applied in a straightforward way to the EF equations. We find an approach that  enables stable propagation long enough to witness non-adiabatic behavior in a model system before non-trivial instabilities take over. Implications for the development and analysis of EF-based methods are discussed.
\end{abstract}

\maketitle  
\section{Introduction}
The exact factorization (EF) of the time-dependent molecular Schr\"odinger equation~\cite{AMG10,AMG12} is an exact reformulation of the quantum dynamics of interacting electronic and nuclear systems. The molecular wavefunction is expressed as a single product of a nuclear wavefunction and an electronic wavefunction that is conditionally dependent on the nuclear coordinate, with two coupled equations describing their motion. Many interesting exact properties of the formalism have been uncovered in recent works, e.g.~\cite{AMG10,AMG12,SAMYG14,KAM15,AASG13, MAKG14, CAG16, FHGS17, RTG16,RPG17, CA17, L15,Curchod_EPJB2018,SASGV17, EA16,SAG16,RG16, LRG18,C15,HARM18,SG17,GZR18},
 shedding light on the nature of interactions between dynamical quantum subsystems as well as on interactions between quantum and classical subsystems beyond adiabatic treatments. From a practical viewpoint, the EF equations provide a rigorous starting point for methods for non-adiabatic dynamics and already we have seen the development of mixed quantum-classical approaches~\cite{AMAG16, MAG15,GAM18,ATC18}, with successful applications in photochemical dynamics~\cite{MATG17}, as well as density functionalizations~\cite{RG16,LRG18}. 

Regarding the exact EF equations, prior studies of the exact features were based on first solving the original full molecular Schr\"odinger equation, and then extracting the exact coupling terms from inverting the exact EF equations. That is, a direct numerical solution of the coupled exact EF equations was avoided. Indeed, the stability and convergence properties of these equations remained unexplored, properties which are of interest when developing further EF-based approximations. In this paper, we discuss unique challenges that a self-consistent numerical solution of the exact coupled EF equations pose, give a numerical solution for a  model problem, and present a preliminary analysis of their stability. We show that the usual numerical  methods developed for the time-dependent Schr\"odinger equation fail  when applied to the EF equations, and investigate these failures with a preliminary formal analysis. We show how we were able to obtain a stable numerical propagation for long enough to witness non-adiabatic behavior in the model system before non-trivial instabilities kill the calculation. 

This paper is structured as follows: section~\ref{sec:EFeqns} briefly reviews the EF formalism. Section~\ref{sec:apriori} then discusses aspects that a numerical solution need to consider, before discussing such a solution for a model problem in Sec.~\ref{sec:SM_sims}. A preliminary analysis of the well-posedness of the EF equations is given in Sec.~\ref{sec:StabA}, before we conclude in Sec.~\ref{sec:concs}. 

\section{The exact factorization equations}
\label{sec:EFeqns}
In the EF approach~\cite{AMG10,AMG12,GG14,H75,H81}, the exact molecular wavefunction is written as a single product of a marginal wavefunction $\chi$ and a conditional wavefunction $\Phi_R$, as
\begin{equation}
\label{eq:ansatz}
\Psi(r,R,t) = \phi_R(r,t) \chi(R,t)
\end{equation}
subject to the partial normalization condition (PNC)
\begin{equation}
\label{eq:PNC}
\int \left| \phi_R(r,t) \right|^2 dr = 1,  \hspace{1em} \forall R, t\;.
\end{equation}
Here, $R$ represent the set of all  nuclear coordinates and $r$ the electronic coordinates.
The factorization above is unique up to phase-transformation that depends only the nuclear coordinates and time:
\begin{align}
\label{eq:gauge}
\phi_R(r,t) &\to e^{i \theta(R,t)} \phi_R(r,t)\hspace{4em}\notag{}\\
\chi(R,t) &\to e^{-i \theta(R,t)} \chi(R,t)\hspace{4em} 
\end{align}
Notice that  Eq.~(\ref{eq:ansatz}) has the same form as the wavefunction in the Born-Oppenheimer approximation, but Eq.~(\ref{eq:ansatz}) is {\it exact}, as proven in the earlier works, and the equations that the two parts $\chi$ and $\Phi_R$ satisfy differ from that in the Born-Oppenheimer equation. These equations may be derived by inserting Eq.~(\ref{eq:ansatz})  into the Dirac-Frenkel action and applying the variational principle~\cite{AMG10,AMG12,AMG13,ACEJ13}, and, for one nuclear degree of freedom and one electronic degree of freedom have the form 
\begin{equation}
\label{eq:elecEqn}
\left[ \hat{H}_{\mathrm{BO}} + \hat{V}^e_{\mathrm{ext}}(r,t)+ \hat{U}\en- \epsilon(R,t)\right]\phi_R(r,t) = i \partial_t \phi_R(r,t),
\end{equation}

\begin{equation} 
\label{eq:nucEqn}
\left[\frac{\left(-i \nabla_R + A(R,t)\right)^2}{2M} + V^n_{\mathrm{ext}}(R,t) + \epsilon(R,t)\right]\chi(R,t)=i \partial_t \chi(R,t),
\end{equation}
where  time-dependent potential energy surface (TDPES) $\epsilon(R,t)$, vector potential $A(R,t)$, and coupling operator $\hat{U}\en$ are given by
\bea
\label{eq:eps}
\epsilon(R,t) &=& \langle \phi_R|\hat{H}_{\mathrm{BO}} + \hat{V}^e_{\mathrm{ext}} + \hat{U}\en-i \partial_t |\phi_R \rangle_r \\
A(R,t) &=& -i \langle \phi_R(r,t) | \nabla_R  \phi_R(r,t) \rangle_r \\
\nonumber
\hat{U}_{\mathrm{en}}[\phi, \chi] &=& \frac{1}{M}\left[ \frac{(-i \nabla_R - A)^2}{2} \right.\\ 
&+&\left.\left (\frac{-i\nabla_R \chi}{\chi}+ A \right)\cdot\left(-i \nabla_R - A\right) \right]
\label{eq:UenPhi}
\eea
with $\langle ... \rangle_r$ indicating an inner product only over the electronic space. The potentials $V\ext^e$ and $V\ext^n$ represent potentials that are externally applied to the coupled system of electrons and nuclei, e.g. a laser field, and $\hat{H}_{\rm BO}$ is the usual Born-Oppenheimer Hamiltonian consisting of electronic kinetic energy, electron-electron, electron-nuclear, and nuclear-nuclear Coulomb potentials.
Note that atomic units are used throughout, and $M$ denotes the effective mass of the nuclear coordinate. Under the phase-transformation of Eq.~(\ref{eq:gauge}), the potentials $\epsilon$ and $A$ transform in a gauge-like way: 
\begin{align}
A(R,t) \to A(R,t) + \nabla_R\theta(R,t) \notag{}\\
\epsilon(R,t) \to \epsilon(R,t) + \partial_t \theta(R,t)
\end{align}
The TDPES can be separated into a gauge-independent (GI) term and a gauge-dependent (GD) term, $\epsilon(R,t)  = \epsilon_{\rm GD}(R,t) + \epsilon_{\rm GI}(R,t) $ where $\epsilon_{\rm GD}(R,t) = \langle\phi_R(t)\vert -i\partial_t \phi_R(t)\rangle$ and $\epsilon_{\rm GI}(R,t)$ are the remaining terms of Eq.~(\ref{eq:eps}). 
The three terms in Eqs.~(\ref{eq:eps})--~(\ref{eq:UenPhi}) capture the entire coupling of the electrons and nuclei; through them, the solution of the electronic equation depends on the nuclear wavefunction, and the solution of the nuclear wavefunction depends on the electronic wavefunction. 
Refs.~\cite{AMG10,AMG12,AASG13,AASMMG15,MAKG14,RTG16,CA17} (for example),  have emphasized the significance of $\epsilon$ and $A$ in yielding the exact nuclear dynamics; in contrast to the Born-Huang expansion, which is also exact, here the exact nuclear dynamics is achieved with a single potential energy surface rather than an infinite number of them, and a single vector potential coupling rather than an infinite number of first and second-order couplings. Wavepacket branching and decoherence are captured, where a central role is played by $\hat{U}\en$.

\section{A Priori Numerical Considerations}
\label{sec:apriori}
Here we make some observations that will inform our attempt to solve Eqs.~(\ref{eq:elecEqn}) -- (\ref{eq:UenPhi}) self-consistently for a model system. 

\subsection{Reality of potentials $\epsilon(R,t)$ and $A(R,t)$}
\label{sec:reality}
The equation for the nuclear wavefunction is of Schr\"odinger-form provided the potentials $A(R,t)$ and $\epsilon(R,t)$ are real. The reality of the potentials ensures that the Hamiltonian in the nuclear equation is Hermitian and consequently that the norm of the nuclear wavefunction is preserved during the time-evolution.  However, the potentials $A(R,t)$ and $\epsilon(R,t)$ are  determined by the solution of the electronic equation, which not only does not have a Schr\"odinger form, but has a Hamiltonian to which notions of hermiticity do not technically apply
(see also Sec.~\ref{sec:normcons}). Nevertheless, norm-conservation in the electronic equation is still a meaningful concept; indeed it is particularly important for the electronic equation, given the role of the PNC of Eq.~(\ref{eq:PNC}) in ensuring that the factorization is unique. Propagation under the exact EF equations does conserve the partial norm~\cite{ACEJ13,AMG13} and in this subsection we show how this directly ensures the scalar and vector potentials that drive the nuclear equation are real. When the equations are discretized in a numerical simulation, additional considerations need to be made.

To see why the partial norm condition of the electronic wavefunction assures us that the scalar and vector potentials in the Schr\"odinger equation for the nuclear wavefunction are real,  we first observe that
\ben
0 \stackrel{\rm PNC}{=} \frac{\partial}{\partial \lambda} \langle \phi_R(t)\vert\phi_R(t)\rangle_r = -\operatorname{Im}  \left\langle \phi_R(t)\vert -i\partial_\lambda\phi_R(t)\right\rangle_r
\label{eq:normcons_real}
\een
where $\lambda$ could represent $R$ or $t$, i.e. a parameter of the electronic wavefunction. So the expectation value on the right-hand-side will be purely real if the PNC is satisfied. 
 Taking $\lambda = R$, the expectation value appearing on the right-hand-side becomes the vector potential $A(R,t)$ and hence PNC condition ensures $A(R,t)$ is real. Taking $\lambda = t$, the expectation value on right-hand-side becomes the gauge-dependent part of the scalar potential, $\epsilon_{\rm GD}(R,t)$, hence the PNC condition ensures $\epsilon_{\rm GD}(R,t)$ is real. The first two terms in the gauge-independent part $\epsilon_{\rm GI}(R,t) = \langle \phi_R|\hat{H}_{\mathrm{BO}} + \hat{V}^e_{\mathrm{ext}} + \hat{U}\en|\phi_R \rangle_r$ 
 are real due to the operators $\hat{H}_{\mathrm{BO}}$ and  $\hat{V}^e_{\mathrm{ext}}$ being Hermitian, while the last term can be shown to be $\langle \phi_R\vert \hat{U}\en\vert\phi_R\rangle_r = \left(\langle \nabla_R\phi_R\vert\nabla_R\phi_R\rangle_r - A^2\right)/2M$ upon making use of the definition of the vector potential to cancel some terms. Thus, we have shown that the potentials $\epsilon(R,t)$ and $A(R,t)$ appearing in the Schr\"odinger equation for the nuclear wavefunction are real, provided that the PNC is satisfied.

However, while the above arguments hold for the exact (continuous) case, there is no guarantee they will always hold for a numerical solution because of discretization. For a given discretization, the analog of the relation Eq.~(\ref{eq:normcons_real}) requires a certain additional choice of the discrete points at which to evaluate the integral. For example, consider the simplest discrete version of Eq.~(\ref{eq:normcons_real}) taking $\lambda = R$: letting $\phi_{iJ}^{(n)}$ represent the conditional electronic wavefunction with $i, J, n$ as the index for $r,R, t$ respectively, 
\bea
\nonumber
&\langle \phi_{R+\Delta R}(t) \vert\phi_{R+\Delta R}(t) \rangle - \langle \phi_{R}(t) \vert\phi_{R}(t)\rangle \to \\
\nonumber
&\frac{1}{\Delta R}\sum_i \vert \phi_{i,J+1}^{(n)}\vert^2 - \vert \phi_{i,J}^{(n)}\vert^2\\\nonumber
&=  
  2\operatorname{Re} \left( \sum_i \frac{(\phi^{(n)*}_{i,J+1}+\phi^{(n)*}_{i,J})}{2}
  \frac{(\phi^{(n)}_{i,J+1}-\phi^{(n)}_{i,J})}{\Delta R} \right)\\
& = -2\operatorname{Im} \left( \sum_i \frac{(\phi^{(n)*}_{i,J+1}+\phi^{(n)*}_{i,J})}{2}
  \frac{-i(\phi^{(n)}_{i,J+1}-\phi^{(n)}_{i,J})}{\Delta R} \right)
  \label{eq:discreteA}
\eea
The term inside the parentheses in the last line corresponds to
$\langle \phi_R(r,t)| -i \nabla_R \phi_R(r,t) \rangle_r = A(R,t)$. 
As in the continuous case the fact that the norm is independent of $R$
ensures that $A(R,t)$ is real, but here only provided one takes the midpoint in $R$-space to evaluate $\langle \phi_R(r,t)|$, i.e. the average $(\phi^{(n)*}_{i,J+1}+\phi^{(n)*}_{i,J})/2$.

One can do the same calculation with time indices:
\begin{multline}
 \langle \phi_{R}(t+\Delta t) \vert\phi_{R}(t+\Delta t) \rangle - \langle \phi_{R}(t) \vert\phi_{R}(t)\rangle \to \\
   \sum_i \left(|\phi^{(n+1)}_{i,J}|^2 - |\phi^{(n)}_{i,J}|^2 \right) / \Delta t \\
  =  
  2\operatorname{Im} \left( -i\sum_i \frac{(\phi^{(n+1)*}_{i,J}+\phi^{(n)*}_{i,J})}{2}
  \frac{(\phi^{(n+1)}_{i,J}-\phi^{(n)}_{i,J})}{\Delta t} \right)
  \label{eq:t_norm_cons}
\end{multline}
Here the term in parenthesis is 
$\langle \phi_R(r,t)| -i \partial_t \phi_R(r,t) \rangle_r = \epsilon_{\rm GD}$. 
Again, as in the continuous case, PNC implies $\operatorname{Im}(\epsilon_{\rm GD}) =0$
provided $\langle \phi_R(r,t)|$ is evaluated precisely at 
the midpoint in time.

In practise, we used higher-order discretizations in both space and time (see shortly) for which  a more complex additional condition would be required to guarantee that satisfaction of the PNC implies reality of the potentials. 
We found, however, that for the time-steps and spatial-grid we used that neglecting this correction did not yield a large error, i.e. that any imaginary part in the potentials was small. A cruder fix that we tried was to simply to redefine the potentials, e.g. $A(R,t) \to \mathrm{Re} A(R,t) \;{\rm and} \nabla_R A(R,t) \to \mathrm{Re} \nabla_R A(R,t)$, but this did not appear to make much difference in the calculations for which we were able to propagate for a significant amount of time.

\subsection{Hermiticity and partial norm conservation}
\label{sec:normcons}
Having discussed the subtleties in ensuring the reality of the potentials in the numerics assuming we have a propagation scheme that conserves the PNC, we now turn to what such a propagation scheme should be. 

As mentioned in Sec.~\ref{sec:reality}, although the equation for the nuclear wavefunction is of Schr\"odinger-form, for which efficient and and accurate propagation schemes have been well-studied,  the equation for the electronic equation is not. The form of the coupling potential $U_{en}[\phi,\chi]$ is unprecedented in the literature: it is non-linear and also has operators which act on the parametric dependence of the conditional electronic wavefunction.
The domain of this operator is outside the electronic Hilbert space associated with a fixed nuclear configuration; rather, it connects electronic Hilbert spaces associated with neighboring nuclear coordinates. As a result it is not possible to define a Hermitian conjugate, a feature that is only compounded by its non-linearity; the notion of operator Hermiticity is not meaningful  for general non-linear operators while a notion of Hermiticity can be rescued for such operators however in the context of expectation values, as in Ref.~\cite{Schwartz97}. 

One of the most salient consequences of Hermiticity of the Hamiltonian in the usual Schr\"odinger equation, is that it guarantees norm-conservation. As mentioned in Sec.~\ref{sec:reality}, non-linear Hamiltonians may still preserve the norm, and  the Hamiltonian in the electronic equation does. This is particularly important for uniqueness of the EF approach, and for ensuring that the coupling potentials appearing in the Schr\"odinger equation for the nuclear wavefunction are real. 

A commonly-used time propagator for evolving the Schr\"odinger equation is the Crank-Nicolson (CN) propagator~\cite{CN96}, which conserves the norm when the Hamiltonian is Hermitian. 
Considering a general discretized wavefunction $\psi^{(n)}_{i}$, with $n,i$ being the time and spatial index respectively, 
CN yields for the projection of the time-evolution  $\langle \psi(t)\vert -i\partial_t\psi(t)\rangle$, 
\begin{multline}
  \sum_i \frac{(\psi^{(n+1)*}_{i}+\psi^{(n)*}_{i})}{2}
  \frac{-i(\psi^{(n+1)}_{i}-\psi^{(n)}_{i})}{\Delta t} \\
  =
  \sum_{i,l} \frac{(\psi^{(n+1)*}_{i}+\psi^{(n)*}_{i})}{2}
  H_{i,l}
  \frac{(\psi^{(n+1)}_{l}+\psi^{(n)}_{l})}{2}
  \label{eq:usualCN}
\end{multline}
which is real for Hermitian Hamiltonians $H$ (often evaluated at the midpoint in time for time-dependent and/or non-linear $H$). 

We now consider the application of CN for the electronic equation. We start by
expanding Eq.~(\ref{eq:elecEqn}) and grouping terms two by two in the following way:
\begin{multline}
\left( \hat{H}_{\mathrm{BO}} + V^e_{\mathrm{ext}} \right) \phi_R(r,t) 
- \langle \phi_R(t)|\hat{H}_{\mathrm{BO}} + V^e_{\mathrm{ext}}  |\phi_R(t) \rangle_r \phi_R(r,t) \\
- \frac{1}{2M}\nabla_R^2 \phi_R(r,t)
+ \langle \phi_R(t)|\frac{1}{2M}\nabla_R^2 |\phi_R(t) \rangle_r \phi_R(r,t) \\
- \frac{1}{M}\frac{\nabla_R \chi(R,t)}{\chi(R,t)} \cdot \nabla_R \phi_R(r,t)
+ i\left[\frac{1}{M}\frac{\nabla_R \chi(R,t)}{\chi(R,t)} \cdot A(R,t) \right] \phi_R(r,t) \\
= i \partial_t \phi_R(r,t)- \langle \phi_R(t)| i \partial_t |\phi_R(t) \rangle_r\phi_R(r,t) \;,
  \label{eq:simple_phi_eq}
\end{multline}
such that each line contains two terms that cancel each other if one takes the expectation value over
the conditional wavefunction. 
Now, we take the inner product with $\langle \phi_R(r,t)|$ to obtain the analog of Eq.~(\ref{eq:usualCN}). 
Then for partial norm conservation one only has to show that 
$\operatorname{Im}(\epsilon_{GD}) = 0$, with the $\epsilon_{GD}$ defined in 
Eq.~(\ref{eq:t_norm_cons}).
But in this case it is more complicated than the usual case in Eq.~(\ref{eq:usualCN}) as the zero imaginary part does not come from Hermiticity of the matrix. Rather, it results from exact cancellation of terms when the equation is projected on $\langle \phi_R(r,t)|$.
For example $
\langle \phi_R(t)|
- \frac{1}{M}\frac{\nabla_R \chi(R,t)}{\chi(R,t)} \cdot \nabla_R | \phi_R(t) \rangle_r
$
should cancel exactly $
\langle \phi_R(t)|
i\left[\frac{1}{M}\frac{\nabla_R \chi(R,t)}{\chi(R,t)} \cdot A(R,t) \right] | \phi_R(t) \rangle_r
$.
This obviously cancels in the continuous case from the definition of $A(R,t)$ but in the discrete case we have to define each
element carefully.
For example, if one defines 
\begin{equation}
  | \phi_R(t) \rangle \equiv  \frac{(\phi^{(n+1)}_{i,J}+\phi^{(n)}_{i,J})}{2}
\end{equation}
to conserve the norm in a Crank-Nicolson fashion then we have to use an alternative 
definition of $A$, that we call $\tilde{A}(R,t)$ here.
We define $\tilde{A}$ as:
\begin{multline}
  \tilde{A}(R,t) \equiv \frac{-i}{{\cal{N}}^2(R,t)} \times\\
  \sum_i 
  \frac{(\phi^{(n+1)}_{i,J}+\phi^{(n)}_{i,J})}{2}
  \frac{(\phi^{(n+1)}_{i,J+1} - \phi^{(n+1)}_{i,J}) + 
    (\phi^{(n)}_{i,J+1} - \phi^{(n)}_{i,J})}{2\Delta R} 
    \label{eq:tildeA}
\end{multline}
with ${\cal{N}}(R,t)$ the partial norm of the time midpoint: 
\begin{equation}
  {\cal{N}}^2(R,t) =  \sum_i
  \frac{(\phi^{(n+1)*}_{i,J}+\phi^{(n)*}_{i,J})}{2}
  \frac{(\phi^{(n+1)}_{i,J}+\phi^{(n)}_{i,J})}{2}
\end{equation}
which is not equal to one.
Choosing this definition does not enforce that $\tilde{A}(R,t)$ is real though
it is equivalent to the definition following Eq.~(\ref{eq:discreteA}) when both $\Delta t$ and $\Delta R$ go to zero.
Using the definition of Eq.~(\ref{eq:tildeA}) and renormalizing the average values
of the form $\langle \phi_R(t)| O | \phi_R(t) \rangle_r$  
that appear in Eq.~(\ref{eq:simple_phi_eq})  by 
${\cal{N}}(R,t)$ would allow $\operatorname{Im}(\epsilon_{GD})$ to be zero and the 
partial norm to be conserved during the propagation.
The main problem with this approach is dependence of many quantities on the next time-step 
with $\phi^{(n+1)}_{i,J} $ which would create a highly non-linear equation to solve
during the propagation.

The analysis above demonstrates the challenge of discretizing the EF equations in a way that remains faithful to the norm-conservation aspects of the exact equations. In an alternative approach, one could consider instead discretizing first, and then factorizing, however  this presents it own challenges in identifying separate equations with clear definitions for the potentials; some preliminary work in this direction is presented in Appendix~\ref{sec:factafterdisc}. 

In our numerical investigations, we use a standard RK45 scheme (see Sec.~\ref{sec:propagator}), with a small enough time-step and grid spacing that errors in partial norm conservation were small over the duration of our simulation.   We found that  enforcing a crude ``norm-conservation correction" in our propagation in the form of 
\ben
\hat{U}\en \to \hat{U}\en - i\mathrm{Im}\langle\Phi_R(t)\vert \hat{U}\en\vert\Phi_R(t)\rangle.
\label{eq:Uencorr}
\een 
generally improved the stability and accuracy of the dynamics. 

\subsection{EF expansion in the Born-Oppenheimer basis}
Generally, a numerical solution of Eqs.~(\ref{eq:elecEqn}) -- (\ref{eq:UenPhi}) requires spatial grids for the nuclear coordinate $R$ and for the electronic coordinate $r$. Since electronic wavefunctions vary on the scale of tenths of an Angstrom while extending over a few Angstroms at least (more when the system evolves to more highly excited states or partially ionizes), the size of the electronic grid can limit the efficiency of the time-propagation. In some cases, the numerics can be made more tractable by using a basis for the electronic equation; the choice of an optimal basis depends on the physics of the problem, but for cases where external perturbations are weak and the system is initially well-described by one or a few BO states, expanding the conditional electronic wavefunction in a truncated BO basis can make the numerical simulation simpler and faster.

The equation of motion for the electronic system then turns into coupled equations for basis coefficients, $C_j(R,t)$, where
\begin{equation}
\phi_R(r,t)  = \sum_{j} C_j(R,t)\phi^{j,BO}_{R} (r)
\end{equation}
with the adiabatic BO states satisfying $\hat{H}_{BO} (R,r)\phi^{j,BO}_{R} (r) = \epsilon^{j}(R) \phi^{j,BO}_{R} (r)$, with $\epsilon^j(R)$ the $j$th BO potential energy surface.
Inserting this into the equation of motion for the conditional electronic wavefunction leads to (see Appendix~\eref{sec:ApB} for details)
\begin{align}
\label{eq:finalEOM}
i\partial_t C_i(R,t) &=  \left( \epsilon_{i}(R)- \epsilon(R,t)\right) C_i  + U^{i}_{en}(R,t)
\end{align}
where: the $i$th-projected $U\en$ is
\bea
\label{eq:UenC}
U^i\en  
 &=& \frac{1}{M_n}\left[ \left( \frac{i \nabla \cdot A - A^2 - \nabla^2 }{2} +\frac{\nabla \chi}{\chi}\left( i A - \nabla \right) \right)C_i  \right. \notag{}\\
 &-& \left.\sum_{j} \left(\frac{1}{2}d^{2}_{ij} + d_{ij}\nabla + \frac{\nabla \chi}{\chi}d_{ij} \right)C_j \right],
\eea
the non-adiabatic coupling terms are 
\begin{align}
\label{eq:NACVdef}
d_{ij} = d_{ij}(R) = \langle \phi_R^i | \nabla_{R}  \phi_R^j\rangle\,,\; d_{ij}^2 = d_{ij}^2(R) = \langle \phi_R^i | \nabla^2_{R}  \phi_R^j\rangle
\end{align}
and the remaining potential terms are
\begin{align}
\nonumber
A(R,t) & =  -i \left[\sum_{i,j} C_{i}^{*} C_{j} d_{ij} + \sum_{i} C_{i}^{*} \nabla_R C_i \right]\\
\epsilon(R,t) & = \sum_{i} (C_{i}^{*}  U^{i}\en + |C_i|^2\epsilon_i )
\label{eq:expepsA}
\end{align}
The time evolution of the full molecular system then is equivalent to self-consistently evolving Eq.~\eref{eq:finalEOM} with the equation for the nuclear wavefunction, Eq.~\eref{eq:nucEqn}.


\subsection{Choice of gauge}
While the EF equations are invariant under the gauge transform Eqn.~\eref{eq:gauge}, the stability properties may not be. It is possible that a `best gauge' exists for the propagation of these equations, but for now we simply choose the gauge such that
\begin{align}
\label{eq:gaugeCond}
\langle \phi_R | \partial_t \phi_R \rangle &= 0 \notag{}\\
\implies \epsilon(R,t) &= \epsilon_{GI}(R,t)
\end{align}
i.e., the gauge-dependent part of the time-dependent potential energy surface is set to zero. 

The gauge condition Eq.~\eref{eq:gaugeCond} is not guaranteed to hold in a numerical solution and we do not enforce it explicitly. We could attempt to impose the gauge condition in a first-order way, replacing
\begin{equation}
\label{eq:gaugefix}
\partial_t C_i(R,t) \to \partial_t C_i(R,t) - \langle\phi_R\vert \partial_t\phi_R\rangle_r  C_i(R,t)
\end{equation}
However we found that this does not appreciably extend the integration time with the propagation method we used ($ < 5$a.u. difference), even if it does ensure the system remains in the correct gauge for longer.

\subsection{Propagation scheme}
\label{sec:propagator}
An important choice for numerical time-evolution is the propagation algorithm. Much depends on this choice and, for general non-linear systems, it is not always clear why one might be more suitable than another. Since the equation for the nuclear wavefunction has the time-dependent Schr\"odinger form, the Crank-Nicolson (CN) propagator is the preferred choice given its stability and norm-conserving properties. As discussed extensively in Sec.~\ref{sec:normcons}, the electronic equation is not of this form and  CN may not be the best choice due to the non-linearity. In our numerical investigations, we use two propagation schemes. For the majority of the presented results we evolve the electronic system with the fourth-order Runge-Kutta (RK45) scheme, in which the $R$ and $t$-dependent terms vary between the internal steps. Being a purely explicit integrator it is known to be only conditionally stable, and so to investigate RK45's robustness to this problem we also test the simplest purely implicit scheme, the backwards Euler (BE) method, as well as with the mixed explicit-implicit CN scheme. 

The inconsistency in the order of accuracy in this comparison ($O(\Delta t^5)$ and $O(\Delta t^2)$ for RK45 and BE/CN respectively) is somewhat intentional in that it allows us to investigate which is more important for a stable norm-conserving solution: accuracy or an implicit solver. For example, an implicit solver may have absolute stability over a large range of parameters but a low order method may expedite its exit from this region, whereas a more accurate method may be able to remain in its island of stability longer. A direct but preliminary analysis of the stability of the EF equations is carried out in Sec.~\ref{sec:StabA}.

For the two implicit propagation schemes, BE and CN, a matrix solver is required: both BE and CN can be cast into the form ${\cal{M}}\cdot v = b$ for solution vector $v$. To avoid constructing matrices in memory, even when sparse, this equation is solved iteratively using the biconjugate gradient stabilized method (BiCGSTAB)~\cite{vanderVorst}. This method assumes that $\cal{M}$ is not a function of $v$, and so requires linearization of the underlying equation. This is done by not updating the functions $A(R,t)$ and $\epsilon(R,t)$ (which are non-linear in the electronic coefficients) between steps of the BiCGSTAB procedure, leaving only the coefficients themselves and their gradients, all linear functions, free to vary.  

\subsection{Spatial discretization}
The derivative operators,  $\nabla_R$ and $\nabla_{R}^{2} $ are represented by five-point central finite-difference stencils, where the corresponding one-sided stencils are used at the boundaries. To prevent oscillations appearing at the tails of the nuclear wavefunction, $\chi$ and its first and second derivatives are set to zero at the boundaries.


\section{Numerical Results}
\label{sec:SM_sims}
To investigate the propagation of the EF equations, we choose
the one-dimensional Shin-Metiu system~\cite{SM95,AASG13,AASMMG15}, which is a  model of proton-coupled electron-transfer with one nuclear and one electronic coordinate. This model system has been instructive for studying different methods for non-adiabatic dynamics; in particular, the potentials that arise in the EF formalism have been  studied and analyzed in this model, uncovering universal features such as steps in the TDPES bridging BO surfaces after a non-adiabatic event~\cite{AASG13,AASMMG15}. In that work, the full TDSE was solved exactly, finding $\Psi(r,R,t)$, and extracting   the vector and scalar potentials $A(R,t)$ and $\epsilon(r,t)$ via inversions. In contrast, here, we endeavor to directly solve the EF equations, with the potentials emerging on the fly during the propagation.

\begin{center}
\begin{figure}[h]
\includegraphics[width=.5\textwidth]{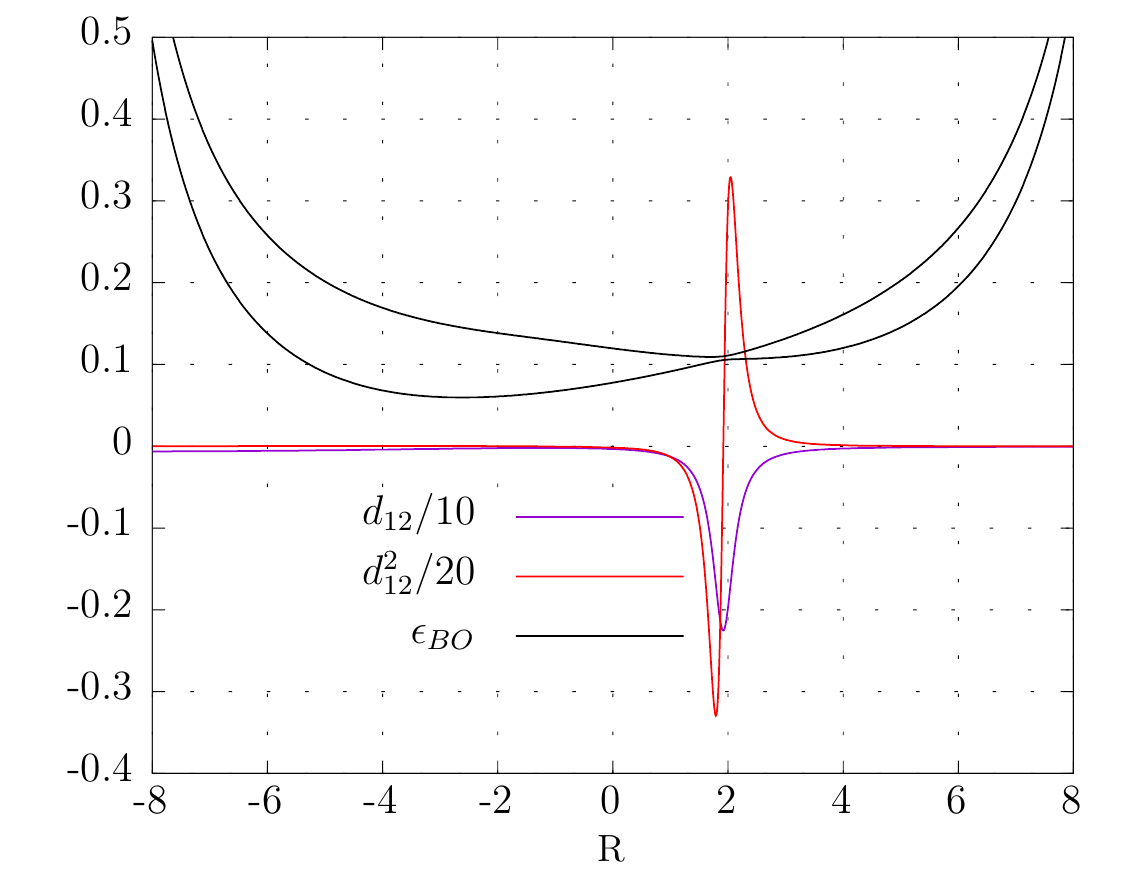}
\caption{
Shin-Metiu model used in all subsequent calculations.  The first two adiabatic surfaces, along with the first and second order non-adiabatic couplings are shown here.}
\label{fig:ShinMetiuFig}
\end{figure}
\end{center}

The Shin-Metiu model Hamiltonian corresponds to an electron and ion moving between two fixed ions a distance $L = 19.0a_0$ apart:
\begin{align}
H(r,R) &=- \frac{1}{2}\partial^{2}_{r}- \frac{1}{2}\partial^{2}_{R} +  \frac{1}{\vert \frac{L}{2}-R\vert}+\frac{1}{\vert \frac{L}{2}+R\vert} \notag{} \\
&-\frac{\mathrm{erf}\left(\frac{\vert R-r\vert}{R_f}\right)}{\vert R-r\vert}
-\frac{\mathrm{erf}\left(\frac{\vert r-\frac{L}{2}\vert}{R_r}\right)}{\vert r-\frac{L}{2}\vert}
-\frac{\mathrm{erf}\left(\frac{\vert r+\frac{L}{2}\vert}{R_l}\right)}{\vert r+\frac{L}{2}\vert}
\end{align}
where the parameters, $M = 1836$ a.u., $R_f = 5.0a_0$, $R_l = 3.1a_0$, $R_r = 4.0 a_0$. Fig.~\ref{fig:ShinMetiuFig} shows the lowest two BO surfaces and the corresponding non-zero non-adiabatic couplings as defined in Eq.~\eref{eq:NACVdef}. The softened interparticle interactions in this model avoids any possible problems with Coulomb singularities~\cite{JSW15}. 

The higher BO surfaces have a large enough energy separation from the first two such that, beginning with occupation only on the lowest two surfaces leads to only these two surfaces ever being occupied, at least for the duration of time that we consider. So we  take a truncated basis of these two adiabatic levels for the electronic system. We take the same initial conditions as used in the earlier work~\cite{AASG13,AASMMG15}, with the system starting on the upper BO electronic surface, $C_1(R,0) = 0$, $C_2(R,0) = 1$ $\forall R$, and with a gaussian nuclear wave function defined by $\chi(R,0) = \exp[-(R-R_0)^2/2\sigma^2]$ with $R_0 = 4$ and $\sigma^2 = 1/5.7$. 

Taking into account the considerations of Sec.~\ref{sec:apriori}, we now embark upon the exact self-consistent solution of the EF equations.
A grid of $2000$ $R$-points is used, corresponding to $\Delta R =0.009$, with a time step of, unless otherwise stated, $\Delta t = 0.1$ a.u. We will compare the results of the EF simulation with the exact solution of the full two-dimensional TDSE with the CN propagator and the same finite difference stencils as above; here we use the same time-step and $R$-grid, and a grid-spacing of $0.3$au for the electronic coordinate. 

A straightforward implementation of the CN propagation for $\chi$ together with the RK45 scheme for $C_i(R,t)$ (including the norm-conservation correction Eq.~(\ref{eq:Uencorr})), unfortunately becomes unstable and unbounded to the point of failure after a very short time, $t_{\mathrm{max}} = 5$a.u., short enough that negligible dynamics have occurred.  When we turned all the coupling terms to zero so that the equations reduced to BO dynamics, the instability vanished, and accurate propagation was obtained for long times, giving identical nuclear and conditional electronic densities as those obtained from the full solution of the electron-nuclear TDSE with zero coupling. This verifies that the problem lies in the coupling terms, and, in particular, the term $\nabla\chi/\chi$ appears to be problematic. 

In the following, we describe different avenues to ``tame" the instabilities, with varying degrees of success.  Considering the best case scenario for different propagators, the maximum times for the simulation before it exploded, were as follows: 370 au for BE, 480 au for CN, and 1490 au for RK45. Below, the results for the RK45 simulation are studied.

\subsection{ Masking the coupling terms}
The instabilities are related to oscillations that initially develop away from the region of appreciable nuclear density before propagating inwards.
 In these regions, since $Re\nabla\chi/\chi = \nabla\vert\chi\vert/\vert\chi\vert$ can become unreliable, and, given that when the nuclear density is very small, the conditional electronic wavefunction has limited physical significance anyway, we apply a mask function which smoothly sets the coupling terms to zero far from the physical region.
 
\begin{center}
\begin{figure}[h]
\includegraphics[width=.5\textwidth]{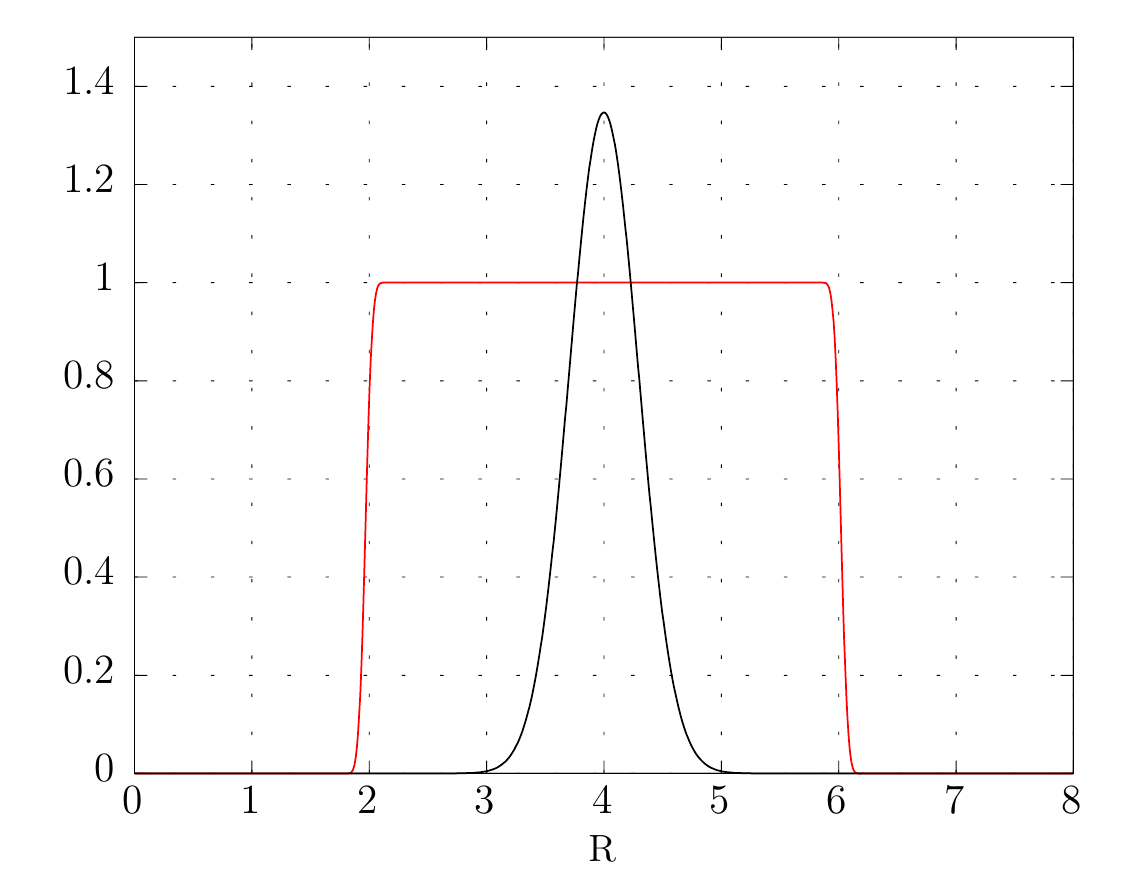}
\caption{
Nuclear density (black) and mask example (red) with tolerance value $\kappa = 10^{-10}$ and steepness  $w=50$ points  (see Appendix~\ref{sec:mask_function}).}
\label{fig:mask_example}
\end{figure}
\end{center}

The function used to generate the mask is defined in Appendix~\ref{sec:mask_function}, and consists of a smooth step on either side of the physical region. The mask `tracks' the density throughout the simulation: we define the center of these steps to be the points at which the nuclear density drops below some threshold $\kappa$ as we approach the physical region from the left and from the  right. 
An example  is depicted in Fig.~\eref{fig:mask_example} along with the density. Unless otherwise specified, the values of $\kappa$ and $w$ indicated in the caption are used.

Outlined in the following subsections\sout{,} are different choices as to which terms to apply the mask and the results on the dynamics.  

\begin{center}
\begin{figure}[h]
\includegraphics[width=0.5\textwidth]{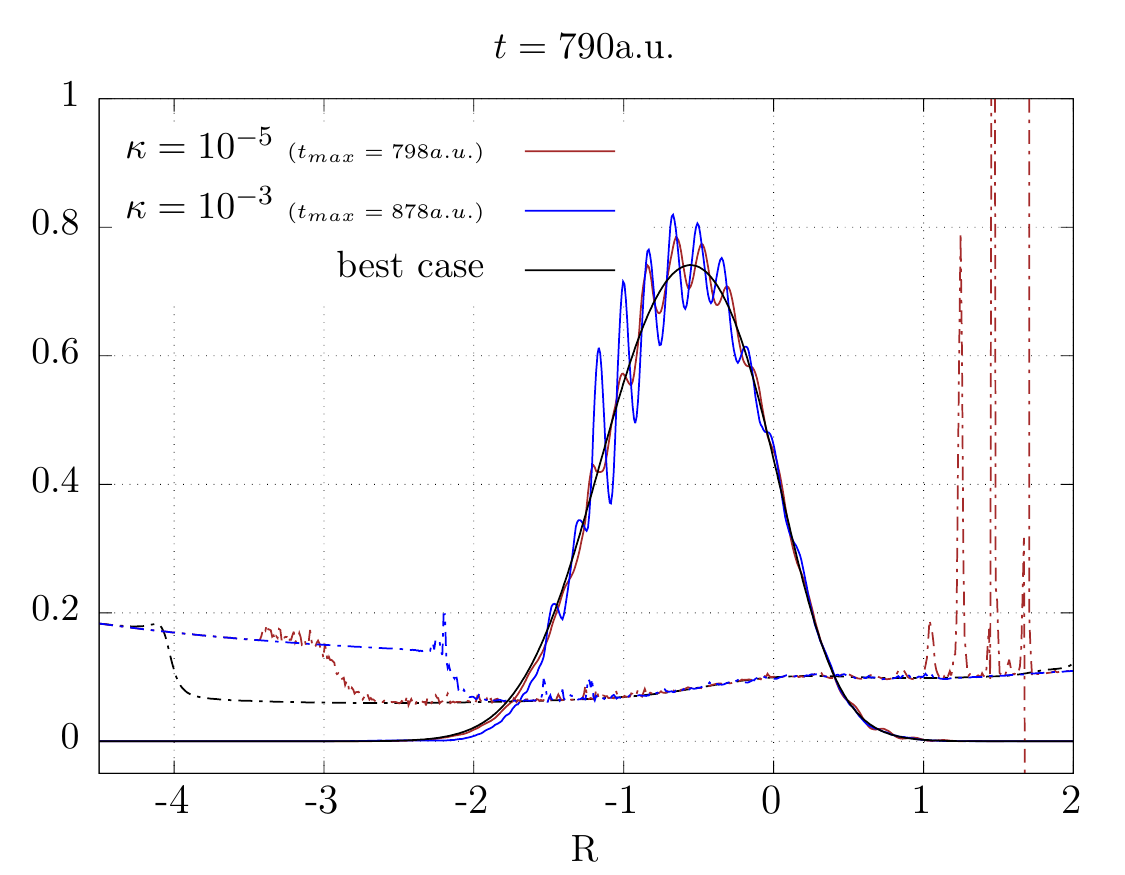}
\caption{
The propagated density (red and blue solid lines) and TDPES  (red and blue dash-dot lines) for the masked $U\en$ simulation at $t = 790$ a.u. for two values of the mask tolerance $\kappa$; the maximum time for their evolutions before the calculation explodes is indicated as $t_{max}$. The black curves indicated by ``best case"  denote the best case scenario in our numerical investigations, and this involves the same mask function as in Fig.~\ref{fig:mask_example} but applied only to $\nabla \chi/\chi$ (see shortly), which gives a density that agrees with the exact solution of the molecular Schr\"odinger equation.  The density in the masked-$U\en$ simulations shown remain smooth up to the time shown, but deviates from the correct solution with large oscillations, and the TDPES and other quantities such as the vector potential (not shown) rapidly develop noise that later cause the simulation to fail. }
\label{fig:mask_uen_grid}
\end{figure}
\end{center}

\subsubsection{Masking $U\en$($t_{\mathrm{max}} = 610$a.u. with $\kappa = 10^{-10}$)} 
Here we apply a mask to the entire coupling term $U\en(r,R)$, so that, for all $r$, at the values of $R$ that have appreciable nuclear density (see above) the conditional electronic equation is as in Eq.~(\ref{eq:elecEqn}) while book-ending either side of this region with purely Born-Oppenheimer dynamics for the conditional electronic wavefunction. The nuclear equation Eq.~(\ref{eq:nucEqn}) is left unaltered. 

With the mask parameters of Fig.~\ref{fig:mask_example},  the calculation fails already at $t_{\mathrm{max}} = 610$a.u. Increasing the value of $\kappa$ allows us to propagate a little further in time, and Figure~\ref{fig:mask_uen_grid} shows the result of propagation with this masked $U\en$, depicting the nuclear density and TDPES at $t = 790$au for two values of the mask edge tolerance $\kappa$. This shows that despite the best efforts of the mask on $U\en$ to keep the influence of noise in the coupling terms in the asymptotic regions at bay, errors still propagate into the region of non-trivial density causing the simulation to fail. 
The level of the rapid oscillations in $\nabla\chi/\chi$ that reaches the area of significant density is apparently not strongly affected by the width of the mask, even for the constrictive case of $\kappa = 10^{-3}$. A mask on $U\en$ alone is is not sufficient to prevent errors from accumulating in the simulation, and  any such errors are not confined to the periphery. 

\subsubsection{Masking $\partial_t C_i(R,t)$: ($t_{\mathrm{max}} =340$a.u.  with $\kappa = 10^{-10}$)}
\begin{center}
\begin{figure}[h]
\includegraphics[width=0.5\textwidth]{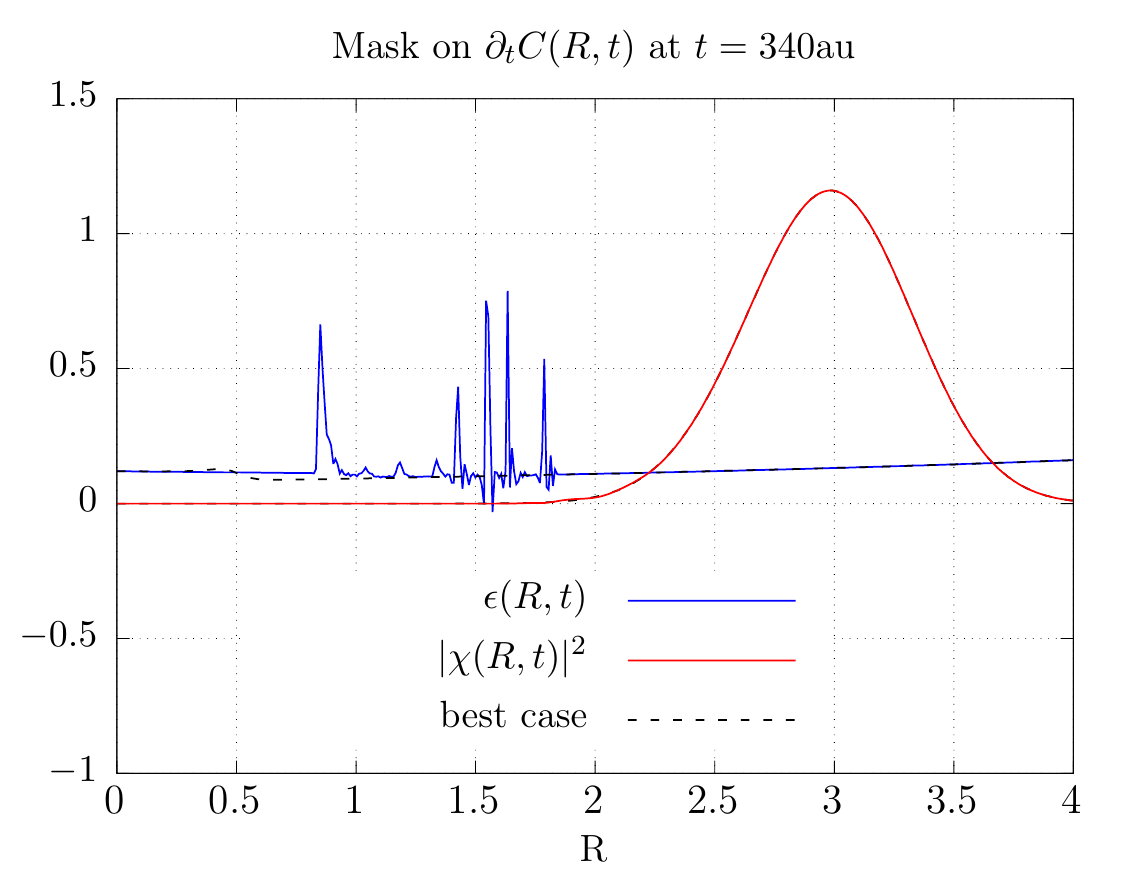}
\caption{
Masked $\partial_t C_i(R,t)$ case at $t=340$a.u showing the density and TDPES a few time-steps before the calculation explodes. This case acquires oscillations in the surface, the vector potential (not shown), and the density  at times before the avoided crossing region is reached with appreciable density.}
\label{fig:mask_dt}
\end{figure}
\end{center}

An alternative way to mask the entire coupling term in the electronic equation is to directly mask the time-derivative of the electronic coefficients, that is, mask the entire right hand side of Eq.~\eref{eq:finalEOM}.
Figure~\ref{fig:mask_dt} shows the result of masking the $\partial_t C_i(R,t)$ functions given in Eqn.~\eref{eq:UenC}. 
Fig.~\ref{fig:mask_dt} shows that this when the mask is placed on $\partial_t C_i(R,t)$ oscillations and spikes exist well inside the masked region even at relatively early times, and the simulation fails before appreciable density has reached the avoided-crossing region. 

\subsubsection{Masking $\nabla\chi/\chi$ and $A(R,t)$ : ($t_{\mathrm{max}} = 1490$a.u. with $\kappa = 10^{-10}$)}
\label{sec:BestCaseScenario}
An alternative to the above two approaches is to mask $\nabla \chi/\chi$ directly, as it is known to be a source of noise and instability early on. 
In fact, applying the same masking procedure as done above greatly improves the stability of the simulation and allows propagation long enough, to a time of $t_{max} = 1220$ a.u., to witness the non-adiabatic event of wavepacket splitting. Additionally, we found that this simulation could be extended further by placing a mask on the vector potential $A(R,t)$ and its divergence $\nabla\cdot A$ at fixed points near boundaries
We use the same mask function as is placed on $\nabla \chi/\chi$, however in this case we fix the start and the end of the mask to be a fixed distance, 100 grid points or 0.9 a.u., from the boundaries. This additional mask on the vector potential did not make a significant difference in the other cases above, but here, masking both $A, \nabla\cdot A$ and $\nabla\chi/\chi$ allowed us to propagate to $t_{max} = 1490$ a.u. We therefore define a moving mask on $\nabla \chi/\chi$ with $\kappa = 10^{-10}$ and $w = 50$, as well as a fixed mask on the edges of the $A(R,t)$ and $\nabla \cdot A(R,t)$ as the \textbf{best case scenario} which we will refer to repeatedly in what follows, and what was plotted in Fig.~\ref{fig:mask_uen_grid} in black.

\subsubsection{Discussion on the best case scenario}
\begin{center}
\begin{figure}[h]
\includegraphics[width=0.5\textwidth]{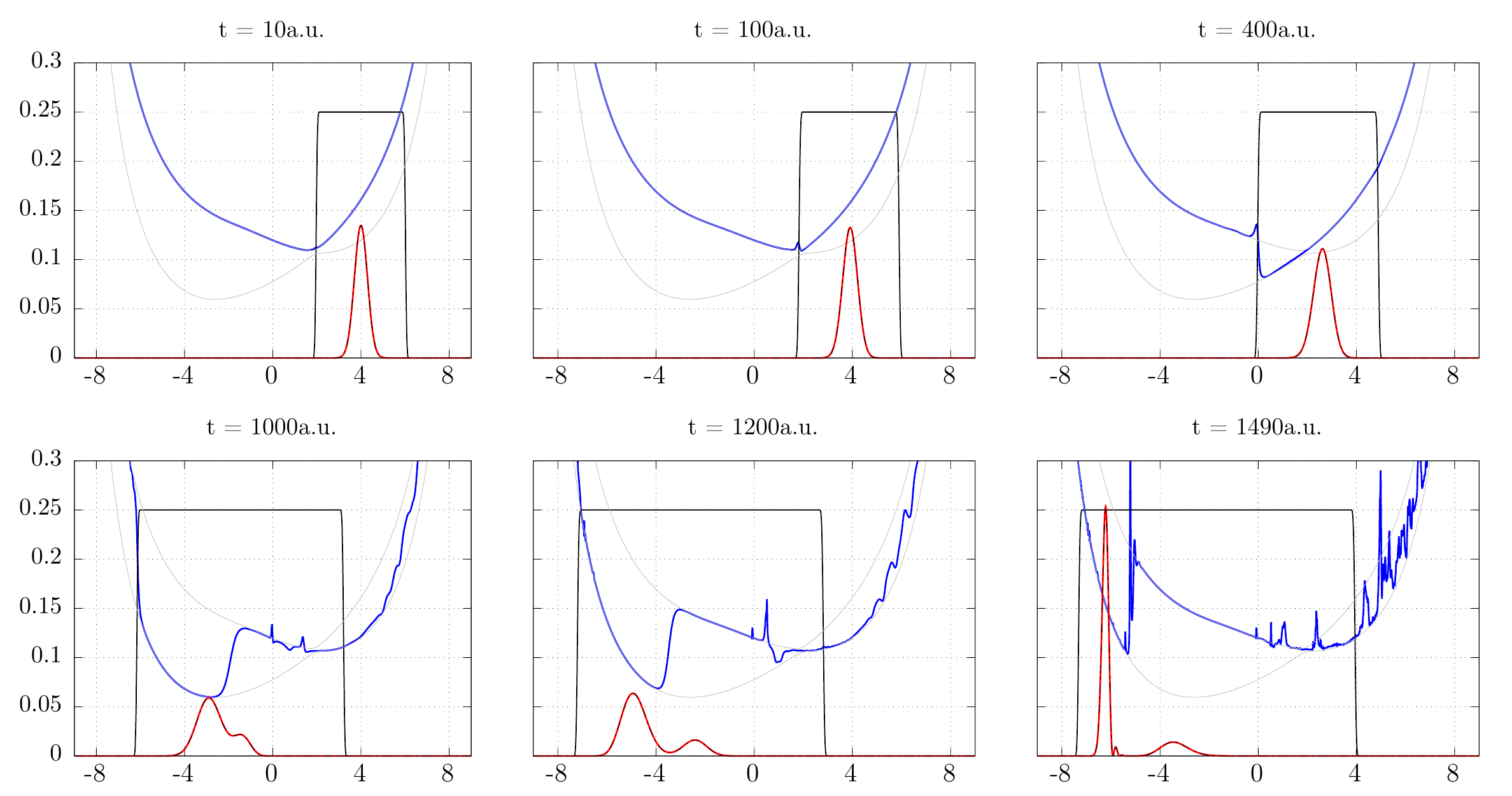}
\caption{ \textbf{Best case scenario:} a moving mask on $\nabla \chi/\chi$ and fixed mask on $A(R,t)$ at the edges with $\kappa = 10^{-10}$ and $w = 50$. Shown are the TDPES (blue) and nuclear density (black) from the simulated EF-equations, along with the exact nuclear density from solution of the full molecular TDSE (red), and BO surfaces (grey). The mask has been scaled by 0.25 and the densities by 1/10 in the figure. Without the mask on $A(R,t)$, the simulation would fail at $1220$a.u., between the final two panels, as the density approaches the boundary.}
\label{fig:mask_gcoc1}
\end{figure}
\end{center}

This solution is depicted in Fig.~\ref{fig:mask_gcoc1} at several times, showing the exact TDSE and EF nuclear densities, the two adiabatic surfaces, the TDPES, and the mask. In all panels the solution agrees well with the direct TDSE solution. One sees the diabatic nature of the TDPES as the density passes through the avoided crossing region, followed by bridging of the two BO surfaces  after passage through the avoided crossing region~\cite{AASG13} that accompanies the splitting of the nuclear wavepacket. We note that in the chosen gauge the gauge-dependent part of the TDPES is zero, so no piecewise off-set is seen~\cite{AASMMG15}.

Noise in the TDPES becomes clearly visible from 1000a.u. onwards, and appears to arise initially where the density is small near the mask boundary; this is where $Re\nabla\chi/\chi = \nabla\vert\chi\vert/\vert\chi\vert$ is largest. We found this also with different values of $\kappa$ which broadens or narrows the mask. 

\begin{center}
\begin{figure}[h]
\includegraphics[width=0.5\textwidth]{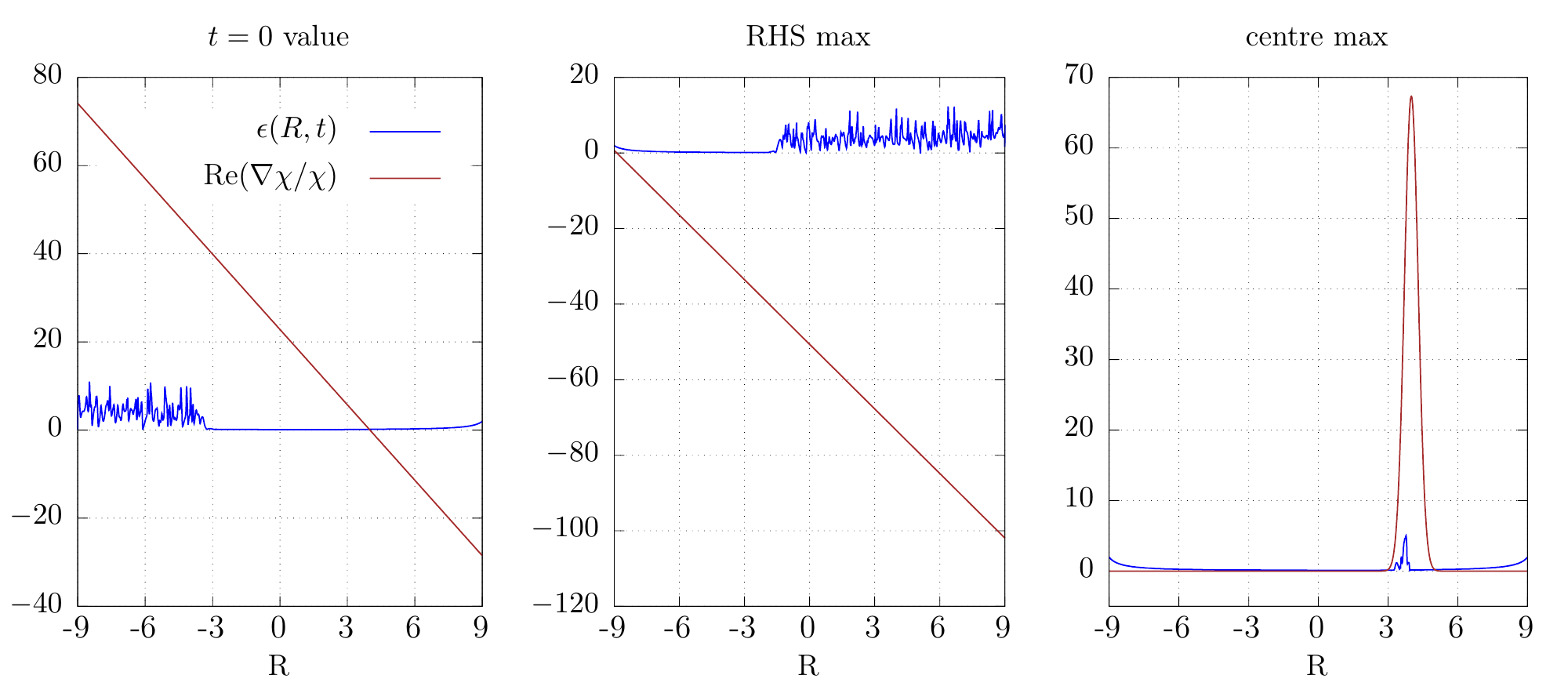}
\caption{Simulation of the electronic system with artificial but analytic  $\nabla \chi/\chi$  functions at $t$ = 10au}
\label{fig:analyticGCOC}
\end{figure}
\end{center}

To investigate whether  simply having large values of $\nabla\chi/\chi$ instigates noise we inserted an analytic form for $\nabla\chi/\chi$ into the electronic equation to see if unstable behavior persists. Naturally, if the term we use bears no relation to the true $\nabla\chi/\chi$ then we are no longer solving the physical problem and any observations we make may have limited scope and utility. Still, it is instructive to understand the numerical properties of this system: with no mask, if  $\nabla\chi/\chi$ is replaced by an analytic form, does the solution of the coupled EF equations still fail quickly? We fix $\nabla\chi/\chi$ to be purely real and have some artificial form. With this we evolve the electronic system, which is now totally decoupled from the nuclear system, and find that
the simulation gives a catastrophically noisy TDPES almost immediately ($<10$au). Fig.~\ref{fig:analyticGCOC} shows simulations, with no masks, for three different forms of $\nabla\chi/\chi$  fed in to the electronic equation: (i) $\nabla\chi(R,t=0)/\chi (R,t=0)$,  (ii) $\nabla\chi(R,t=0)/\chi (R,t=0) + C$ so that it has the same slope but is vertically shifted with $C$ such that the maximum magnitude is on the right-hand boundary, and (iii) a Gaussian with a maximum at the peak density,  which has the least physical resemblance to the actual function. From the results, we immediately observe that one can induce instability in the electronic system at a particular $R$ simply by having $\nabla\chi/\chi$ have a large  value there.   Thus this numerical exploration strongly suggests that a key source of instability in propagation of the exact EF equations is large $\nabla\chi/\chi$. We come back to this point in the preliminary mathematical analysis of the equations in Sec.~\ref{sec:StabA}.

\subsection{Reducing the time step}
\begin{center}
\begin{figure}[h]
\includegraphics[width=0.5\textwidth]{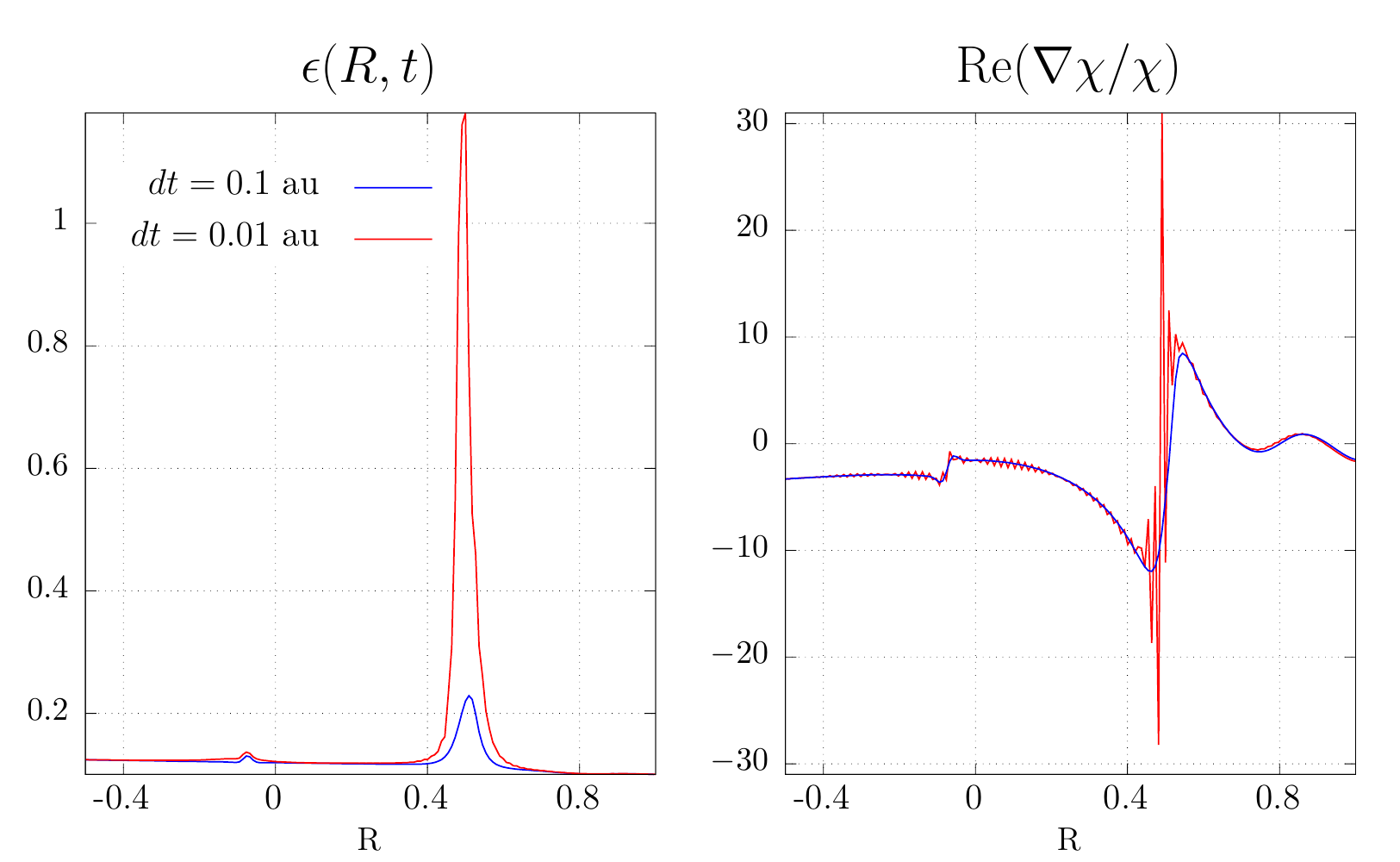}
\caption{ \textbf{Best case scenario:} a moving mask on $\nabla \chi/\chi$ and fixed mask on $A(R,t)$ at the edges. Shown is the time-dependent potential energy surface (left) and Re($\nabla \chi/\chi$) (right) for $\Delta t = 0.1$ (blue) and $\Delta t = 0.01$ (red) at $t=1170$a.u. This illustrates the break down of the smaller-dt solution in a region where the larger dt was well behaved. Note that at this time, the density is located in the region $R \sim  -2$ to $-6$ au which is far from the noise; the point here is that reducing the time step introduces a new instability that was absent with the larger timestep.}
\label{fig:maskgcoc_smallDt}
\end{figure}
\end{center}

A common cause of noise and instability in numerical simulation is violation of a Courant-Friedrichs-Lewy (CFL) condition which provides an upper bound on the time step for a given grid spacing~\cite{CFL}. To investigate this we reduced the step size by an order of magnitude for the ``best case scenario" simulation with the masked $\nabla\chi/\chi$ and $A$. Surprisingly, this simulation failed {\it earlier}, at 1170a.u. The TDPES and Re$(\nabla\chi/\chi)$ at this time are plotted for both time-steps in Fig.~\ref{fig:maskgcoc_smallDt}, showing that indeed the smaller $\Delta t$ choice develops oscillations and sharp features where the larger $\Delta t$ choice has none. A more detailed analysis of this situation is given in Sec.~\eref{sec:StabA}.  

\section{ Stability analysis }
\label{sec:StabA}
The fact that reducing the time-step causes the numerical solution to become unstable at shorter times indicates that the stability properties of the system of equations, Eqs.~(\ref{eq:elecEqn}) -- (\ref{eq:UenPhi}), are unusual and highly non-trivial. The electronic equation has an unprecedented form, whose mathematical and numerical properties are unknown. Do numerical methods exist in which the solution is stable and converged to the true solution? Here we only begin to scratch the surface of the question of whether, once discretized in time and space, the equations are numerically stable, with standard propagation methods. We do not address here the separate question of whether the exact EF equations are mathematically well-posed. 

The nuclear equation, Eq~(\ref{eq:nucEqn}), is a time-dependent Schr\"odinger equation for which stability properties are well-known for potentials that are possibly time-dependent but pre-determined (i.e. not determined by a self-consistent coupled solution as in the EF case). 
Specifically, we consider the equations for the electronic coefficients, $C_i(R,t)$, which we place into a vector, $(C_1(R,t), C_2(R,t)...C_k(R,t))$
($k$ truncated to 2 in our model system of Sec~\ref{sec:SM_sims} but now we keep it general). 
Then we write Eq.~(\ref{eq:finalEOM}) as a matrix equation, {\it as if it was a linear} equation,
\begin{align}
\label{eq:Ham}
i\partial_t \vec{C} &= \left( \underline{f}_0 + \underline{f}_1 + \underline{g}\right)\vec{C}\notag{}\\
&= \underline{K}\vec{C}
\end{align}
where underline denotes a matrix, and we have separated out the terms as follows:
\newline
$\underline{f}_0$ has a Schr\"odinger structure, i.e. matrix proportional to $\delta_{ij}$  with elements $\epsilon^i(\bR,t) - \epsilon(\bR,t) - \nabla^2/2M$. This is diagonal in sense of not coupling different $C_i$, but it is not diagonal once $C_i$ is resolved in $R$ due to the Laplacian, i.e. it would be block-diagonal when writing each coefficient out on a grid $(C_1(R_1,t), C_1(R_2,t)...C_1(R_N,t), C_2(R_1,t), C_2(R_2,t).....)$
\newline
 $\underline{f_1}$ contains all the other purely multiplicative terms (including those coupling the different $C_i$'s via $d_{ij}$ and $d^{2}_{ij}$).
 \newline
 $\underline{g} $ contains terms involving the first derivative $\nabla_R$ and is \textit{not} diagonal in $C_i$. 

The actual equations for the coefficients are however not linear  since $A$, $\epsilon$ depend on $C_i$ (see Eq.~(\ref{eq:expepsA})), but to simplify the analysis for a preliminary examination of stability, we consider these as just some functions of $(R,t)$, and further, that they are determined ahead of time, i.e. they are not produced by some iteration with the nuclear equation, as they would be in actuality. 
With this (big) simplification, what can we say about stability of numerical solutions of this system?

Discretizing in time, we write Eqs.~(\ref{eq:Ham}) as
\begin{equation}
 \vec{C}^{n+1} = \underline{G}\vec{C}^{n}
\end{equation}
where $n = 0,1,2..$ labels the time-step. The system will be numerically stable if the magnitude of the eigenvalues of the amplification matrix G are bounded from below by 1. 
We will consider the simplest schemes: forward and backward Euler. From experience with the TDSE and the $\underline{f}_0$ ``backbone" of our equation, we do not expect that forward Euler will be stable, but it is nevertheless instructive to recall why.  

The forward Euler (FE) method is defined by
\begin{equation}
\label{eq:FE}
i \frac{\vec{C}^{n+1}- \vec{C}^{n} }{\Delta t} = \underline{K}\vec{C}^{n}
\end{equation}
and so
\begin{equation}
 \underline{G}_{\mathrm{FE}} = \left( \underline{1} -i \Delta t \underline{K}\right)
\end{equation}

If we take just the TDSE-type terms, $\underline{f}_0$, this is unconditionally unstable for any finite time-step, because then $\underline{K} = \underline{f}_0$, being Hermitian, has real eigenvalues, which leads to the (complex) eigenvalues of $\underline{G}_{\mathrm{FE}}$, $\lambda_{\rm FE} =1 - i \Delta t \lambda_K$ where $\lambda_K$ are the (real) eigenvalues of $\underline{K}$, having magnitude $\vert\lambda_{\rm FE}\vert = \sqrt{1 + \Delta t^2 \lambda_K^2}$, greater than 1. 
 With all the terms included in $\underline{K}$, it will also not be stable in general, especially given that  at the early times the dynamics occurs on the first excited BO surface, with $\underline{f}_0$ dominating $\underline{K}$, until appreciable density approaches the avoided crossing. 

We next consider an implicit method, as is usually done for TDSE, the simplest of which is the backward Euler (BE) method:
\begin{align}
\label{eq:BE}
 i\frac{\vec{C}^{n+1}- \vec{C}^{n} }{\Delta t} &= \underline{K}\vec{C}^{n+1} \implies   \underline{G}_{\mathrm{BE}} = \left( \underline{1} + i \Delta t \underline{K}\right)^{-1}
\end{align}
Noting that eigenvalues of the matrix inverse are the inverse of the eigenvalues of the matrix, we see that for $\underline{K} = \underline{f}_0$, the eigenvalues have magnitude $\vert\lambda_{\rm 0,BE}\vert = \frac{1}{\sqrt{1 + \Delta t^2 \lambda_K^2}}\approx 1 - \Delta t^2 \lambda_K^2/2$ which is less than one for any non-zero $\Delta t $, so this is unconditionally stable, as is well known for TDSE-form equations. For the remainder of this section, we consider only BE. 

Now we ask how the situation changes when the coupling terms in $\underline{f}_1$ and $\underline{g}$ are included. Because these terms are not Hermitian, the eigenvalues of $\underline{K}$  are no longer guaranteed to be real. Their imaginary part, once multiplied by $i\Delta t$, yields a correction to the real part of the eigenvalue of $G_{\rm BE}$, decreasing or increasing it away from 1, as well as contributing to the imaginary part. 
To make the analysis tractable, we adopt  two simplifications. First, we truncate the number of coefficients to two (as was relevant in the numerical example of the previous section). 
Second, we use a minimal spatial grid for the finite-difference derivatives, taking a 
three-point centered stencil for the Laplacian and a two-point centered stencil for the first derivative. Third, we consider all the other $R$-dependent functions ($\epsilon$, $\epsilon_i$,$d_{ij}$, $d^{2}_{ij}$) as spatially-uniform; which is not a bad approximation within the 3$\times$3 spatial-grid truncation if our $R$-grid is fine enough. Although these approximations may seem drastic, they are valuable in that they straightforwardly lead to a preliminary analysis of the stability properties of the electronic equation. With these three approximations in hand, we write
\begin{equation}
\underline{G}_{\rm BE}^{-1} = \underline{1}+ i \Delta t \underline{K} =  \underline{1} + i \Delta t\begin{bmatrix}
   A_1 & B_{1,2}\\
   B_{2,1} & A_2
\end{bmatrix} 
\label{eq:Gmatrix}
\end{equation}
where $A_i$ and $B_{i,j}$ are tridiagonal matrices in the basis $\left[ n-1, n, n+1\right]$ with elements
\begin{align}
A_i &= \left[ A_{i}^{-1}, A_{i}^{0}, A_{i}^{1}\right],\notag{} \\
B_{i,j} &= \left[ B_{i,j}^{-1}, B_{i,j}^{0}, B_{i,j}^{1}\right],\notag{} \\
\end{align}
where 
\bea
A_{i}^{-1} &= &\frac{1}{2M \Delta x}\left( \frac{\nabla \chi}{\chi} - \frac{1}{\Delta x}\right) ,\notag{}\\
A_{i}^{0} &=&\frac{1}{M}\left( \frac{1}{\Delta x^2} - \frac{A^2-i \nabla\cdot A-2i A \nabla \chi/\chi}{2} \right) - \epsilon+\epsilon_i-\frac{d^{2}_{i,i}}{2} ,\notag{} \\
A_{i}^{1} &=& \frac{1}{2M \Delta x}\left( -\frac{\nabla \chi}{\chi} - \frac{1}{\Delta x}\right) 
\eea
and 
\begin{align}
B_{i,j}^{-1} &= \frac{d_{i,j}}{2\Delta x} =-B_{i,j}^{1},\notag{} \\
B_{i,j}^{0} &= -d_{i,j}\frac{\nabla \chi}{\chi} - d^{2}_{i,j}/2
\end{align}
For small $\Delta t$, we find only one non-trivial eigenvalue, which in this small-$\Delta t$ limit, is 
\begin{align}
|\lambda_{\rm BE}| \simeq  1 + \frac{6\Delta t}{M} \left(\frac{\nabla\cdot A}{2} + A \frac{\nabla \vert\chi\vert}{\vert\chi\vert}\right)  + O(\Delta t^2)
\end{align}
If this is greater than 1 for some $(R,t)$ then the solution locally is exponentially increasing, leading to an instability; a similar result holds for the Forward Euler. The local stability, for small time-steps $\Delta t$, depends then in a subtle way on the sign and size of 
\ben
\frac{\nabla\cdot A(R,t)}{2} + A(R,t) \frac{\nabla \vert\chi(R,t)\vert}{\vert\chi(R,t)\vert}
\label{eq:lam_corr}
\een
 and if we were to try to define a CFL-condition, it would be a spatially- and time- dependent condition, depending on the self-consistent solution of $\chi$ and $\phi$ at each time-step and spatial $R$-point. This result relates back to our numerical observation that instability appears to be triggered when $\nabla\vert\chi(R,t)\vert/\vert\chi(R,t)\vert$ gets large in magnitude. If $\Delta t$ is chosen small enough, it is necessary at least for the term in Eq.~(\ref{eq:lam_corr}) to be negative, but there is no 
 obvious physical reason why these  terms should be negative everywhere.

The analysis above relied heavily on simplifications, exploring only the simplest finite-difference and time-propagation schemes. It is quite possible that a more sophisticated numerical scheme can be developed that is unconditionally stable. The results of the analysis should therefore not be a deterrent to finding such a scheme, but rather serve to point out that the stability conditions are far more complex than in most time-propagation schemes we have encountered.


\section{Conclusions and Outlook}
\label{sec:concs}
The potentials and coupling operator in the EF equations have already provided much insight into fundamental aspects of coupled dynamics of subsystems, while a mixed quantum-classical scheme based on them has already proven to be a soundly-based approach for practical problems. The investigations here on the self-consistent numerical solution of the equations have fundamental interest but also a practical interest: it would be very useful in developing mixed quantum-classical approximations to test the effect of approximations made to the terms prior to making any classical or semiclassical approximation. However, this work has shown such an endeavor is very challenging. Care must be taken to ensure that features as fundamental as having real potentials and partial norm conservation are generally robust in the numerical scheme, yet the structure of the electronic equation shows the usual propagation methods will not respect this. 
These methods applied to propagate the EF equations become unstable rather quickly. 
We showed that the numerical instability can be tamed by using mask functions on the coupling term $\nabla\chi/\chi$, and the vector potential $A$, such that numerically-exact propagation of the EF equations in a model system can be achieved for a time long enough that non-adiabatic branching phenomena and wavepacket splitting can be observed. But even with the masks, instabilities kill the calculation soon after. 
The purpose of this work was not to present an exhaustive exploration of different algorithms but rather to illustrate some of the unforeseen challenges when attempting a direct solution. One cannot pinpoint one term that is the culprit causing the instability, as we saw instabilities remain when $\nabla\chi/\chi$ was replaced with a smooth, analytic, function. That  there is a complex interplay of terms is also suggested by our preliminary stability analysis exploring the CFL-condition for an implicit propagation scheme.
We hope that this paper will motivate further work to take these developments further for a fuller understanding of the stability of the EF equations and approximations. 

\acknowledgments
 We thank Eric Cances for helpful discussions. 
We are grateful to  the U.S. National Science Foundation grant  CHE-1566197 (L. L. and  G. G.), Department of Energy Office of Basic Energy Sciences, Division of Chemical Sciences, Geosciences and Biosciences grant DE-SC0015344 (N. T. M. and G. G. ), and the Research Corporation for Science Advancement Cottrell Scholar Seed Award (G.G.) for support of this work.

\appendix

\section{Factorization Following Discretization}
\label{sec:factafterdisc}
We saw in Sec.~\ref{sec:normcons} that discretization of the EF equations raises challenging issues for conventional time-propagators. 
Instead of factorizing and then discretizing, one can choose a certain discrete propagator for the full wavefunction first, and then factorize the discrete equation. 
The two equations obtained this way should be numerically equivalent to 
the first equation.

Here we start with the equation for $\Psi^{(n+1)}_{i,j}$ in the Crank-Nicolson scheme:
\begin{equation}
  i\frac{\Psi^{(n+1)} - \Psi^{(n)}}{\Delta t} 
  =
  (\hat{H}_{BO} + \hat{T}_n
  + \hat{V}^n_{\mathrm{ext}} 
  + \hat{V}^e_{\mathrm{ext}} )\frac{\Psi^{(n+1)} + \Psi^{(n)}}{2}
  \;.
  \label{eq:cn_psi}
\end{equation}
We define the following compact notations: 
\begin{gather*}
  \nabla^{-1}_R \phi^{(n)}_{i,j}  =  \frac{\phi^{(n)}_{i,j} - \phi^{(n)}_{i,j-1}}{\Delta R} 
  \\
  \nabla^{+1}_R \phi^{(n)}_{i,j}  =  \frac{\phi^{(n)}_{i,j+1} - \phi^{(n)}_{i,j}}{\Delta R}
  \\
  \nabla^{0}_R \phi^{(n)}_{i,j}  =  \frac{\phi^{(n)}_{i,j+1} - \phi^{(n)}_{i,j-1}}{2\Delta R}
  \\
  \partial_t \phi^{(n)}_{i,j} = \frac{\phi^{(n+1)}_{i,j} -\phi^{(n)}_{i,j}}{\Delta t}
  \;.
\end{gather*}
Now we write $\Psi^{(n)}_{i,j} = \chi^{(n)}_{j}\phi^{(n)}_{i,j}$ and expand the operators 
$\partial_t$ and $\hat{T}_n$ for the discretized wavefunctions. 
We write here the first and second order derivatives of a product of two arbitrary functions $g$ and $f$:
\begin{equation}
  \begin{split}
    &\frac{f_{i+1}g_{i+1} - f_{i}g_{i}}{\Delta x}
    = \frac{f_{i+1}-f{i}}{\Delta x}g_{i} + f_{i+1}\frac{g_{i+1}-g{i}}{\Delta x} \\
    &= \frac{(f_{i+1}-f_{i})}{\Delta x}\frac{(g_{i+1}+g_{i})}{2} 
    + \frac{(f_{i+1}+f_{i})}{2}\frac{(g_{i+1}-g_{i})}{\Delta x}
  \end{split}
\end{equation}

\begin{equation}
  \begin{split}
    &\frac{g_{i+1}f_{i+1}+g_{i-1}f_{i-1}-2g_{i}f_{i}}{(\Delta x)^2} 
    \\ 
    & = g_{i}\frac{(f_{i+1}+f_{i-1}-2f_{i})}{(\Delta x)^2} 
    + \frac{(g_{i+1}-g_{i})}{\Delta x} \frac{(f_{i+1}-f_{i})}{\Delta x} \\
    & + \frac{(g_{i}-g_{i-1})}{\Delta x} \frac{(f_{i}-f_{i-1})}{\Delta x}
    + \frac{(g_{i+1}+g_{i-1}-2g_{i})}{(\Delta x)^2} f_{i}
  \end{split}
\end{equation}
which can be written in a more compact way as
\begin{align}
  &\nabla^{+1}fg = (\nabla^{+1}f) g_{+1/2} + f_{+1/2}(\nabla^{+1}g) \\
  &\nabla^{-1}fg = (\nabla^{-1}f) g_{-1/2} + f_{-1/2}(\nabla^{-1}g)
\end{align}
and
\begin{equation}
  \Delta gf = g\Delta f + \nabla^{+1}g\nabla^{+1}f 
  + \nabla^{-1}g\nabla^{-1}f + \Delta g f
  \;.
\end{equation}
We see the notion of midpoints (Sec.~\ref{sec:reality}) appear again.

Using these equations we rewrite Eq.~(\ref{eq:cn_psi}):
\begin{equation}
  \begin{split}
    &(i\partial_t \chi^{(n)}) \phi^{(n+1/2)} +  \chi^{(n+1/2)} (i\partial_t \phi^{(n)})
    \\
    &=
    (\hat{H}_{BO} 
    + \hat{V}^n_{\mathrm{ext}} 
    + \hat{V}^e_{\mathrm{ext}})\frac{\Psi^{(n+1)} + \Psi^{(n)}}{2} \\
    &+ \frac{1}{2M} (\phi^{(n)}\Delta_R\chi^{(n)} 
    + \nabla_R^{+1}\chi^{(n)}\nabla_R^{+1}\phi^{(n)}
    + \nabla_R^{-1}\chi^{(n)}\nabla_R^{-1}\phi^{(n)}
    )\\
    &+ \frac{1}{2M} (\phi^{(n+1)}\Delta_R\chi^{(n+1)} 
    \\
    &\qquad + \nabla_R^{+1}\chi^{(n+1)}\nabla_R^{+1}\phi^{(n+1)}
    + \nabla_R^{-1}\chi^{(n+1)}\nabla_R^{-1}\phi^{(n+1)}
    )
  \end{split}
  \label{eq:discrete_eqmotion}
\end{equation}
where $\phi^{(n+1/2)} = \frac{\phi^{(n+1)}_{i,j} + \phi^{(n)}_{i,j}}{2} $.
In the continuous case one would take the inner product with $\langle \phi_R(r,t)|$ to
obtain the equation of motion for $\chi(R,t)$. In this case, the equivalent would be 
to multply Eq.~(\ref{eq:discrete_eqmotion}) by $\phi^{(n+1/2)}/\cal{N}$ and integrate (sum)
over $i$. Contrary to the coninuous case, in the discrete case the right-hand-side 
does not simplify and identifying quantities like $A(R,t)$ or rewriting in a familiar 
form with a $\hat{U}_{en}$ seems impossible.

 \section{Equation of motion for $C_j (R,t)$}
 \label{sec:ApB}
 
 In the factorization $\Psi(r,R,t) = \phi_R(r,t)\chi(R,t)$ the equation of motion for $\phi_R(r,t)$ is given by
\begin{equation}
\left(\hat{H}_{el} - \eps(R,t) \right)\phi_R(r,t) = i \partial_t \phi_R(r,t)
\end{equation}
with electronic hamiltonian given by
\begin{equation}
\hat{H}_{el} = \hat{H}_{BO}(r,R) + \hat{V}^{e}_{ext}(r,R) + \hat{U}_{en}[\phi_R, \chi](r,R,t)\,.
\end{equation}
In the following calculations for simplicity we neglect the external potential as it does not enter into the Shin-Metiu test case currently under consideration. 
Rearranging terms in Eq.~\ref{eq:UenPhi}, we write the coupling potential as
\bea
\nonumber
\hat{U}_{en} &=& \frac{1}{M}\left( \frac{-\nabla^{2}_{R} +i \nabla_R\cdot A(R,t)- A(R,t)^2}{2} \right.\\
&+&\left. \frac{\nabla_R \chi(R,t)}{\chi(R,t)} \left( i A(R,t) - \nabla_R \right)\right)
\label{eq:uen}
\eea
where $A(R,t)$ is the vector potential given by $A(R,t) = \langle \phi_R(r,t) | -i \nabla_R \phi_R(r,t) \rangle$. 

We now expand the electronic wavefunction in terms of Born-Oppenheimer (BO) basis vectors as 
\begin{equation}
\phi_R(r,t)  = \sum_{j} C_j(R,t)\phi^{j,BO}_{R} (r)
\end{equation}
where
\begin{equation}
\hat{H}_{BO} (R,r)\phi^{j,BO}_{R} (r) = \eps^{j}(R) \phi^{j,BO}_{R} (r)
\label{eq:BOeq}
\end{equation}
with $\eps^j(R)$ the $j$th BO surface. 
All spatial derivatives of $\phi_R(r,t)$ transform into spatial-derivatives of  the basis, which, being time-independent, need only be computed once at the outset of the simulation, and are thus immune to error propagation over the course of the simulation.

For brevity we now drop the `BO' superscript for the basis vectors and so henceforth all wavefunctions denoted by $\phi^{j}_{R}(r)$ are assumed to be BO vectors.

Projecting the electronic equation,
\begin{equation}
\left(\hat{H}_{BO} +\hat{U}_{en} - \eps(R,t)  \right) \sum_{j} C_j(R,t)\phi^{j}_{R} (r) = i \partial_t \sum_{j} C_j(R,t)\phi^{j}_{R} (r)\;,
\end{equation}
onto the state $\langle \phi^{i}_{R}(r)|$ we obtain
\begin{widetext}
\begin{align}
    \sum_{j}  \langle \phi^{i}_{R}(r)| \hat{U}_{en} \left[ C_j(R,t)|\phi^{j}_{R} (r) \rangle \right ] + \left[ \eps_{i}(R) - \eps(R,t) \right] C_i(R,t)  & =  i  \sum_{j} \langle \phi^{i}_R(r)| \phi^{j}_{R}(r)\rangle \partial_t C_{j}(R,t) = i  \partial_t C_{i}(R,t)
 \end{align}
 where we have used Eq.~(\ref{eq:BOeq}) and orthogonality of the basis vectors to collapse a number of sums. 
 
To compute $ \sum_{j}  \langle \phi^{i}_{R}(r)| \hat{U}_{en} \left[ C_j(R,t)|\phi^{j}_{R} (r) \rangle \right ] $ we
expand the derivatives of Eq.~(\ref{eq:uen}):
\begin{align}
\hat{U}_{en} \left[ C_j(R,t) \phi^{j}_{R}(r) \right] = \frac{1}{M}\left[ \frac{-\left( C_j \nabla^2 \phi^j + \phi^j \nabla^2 C_j + 2 (\nabla C_j) (\nabla \phi^j) \right)  + \left ( i \nabla \cdot A - A^2\right) C_j\phi^j}{2} 
 + \frac{\nabla \chi}{\chi} \left(iA C_j\phi^j - (C_j\nabla \phi^j + \phi^j \nabla C_j) \right)\right] 
\end{align}
Projecting onto state $\langle \phi^{i}_{R}(r)|$ we get
\bea
U\en^{i}\equiv \sum_{j} \langle \phi^{i}_{R}(r)|\hat{U}_{en} \left[ C_j(R,t) |\phi^{j}_{R}(r)\rangle \right] =   \frac{1}{M} \left[ -\frac{1}{2}\sum_{j} C_j \langle \phi^i | \nabla^2 | \phi^j\rangle - \frac{1}{2}\nabla^2 C_i \right. \\  - \left.\sum_{j} (\nabla C_j) \langle \phi^i | \nabla | \phi^j \rangle  + \left(\frac{i \nabla \cdot A - A^2}{2}+ \frac{\nabla \chi}{\chi} i A\right)C_i - \frac{\nabla \chi}{\chi}\sum_{j} C_j \langle \phi^i | \nabla|\phi^j \rangle - \frac{\nabla \chi}{\chi}\nabla C_i\right] 
\eea
\end{widetext}
Defining the non-adiabatic coupling vectors as in Eq.~(\ref{eq:NACVdef}), we find Eq.~(\ref{eq:UenC}). 

Likewise, expanding the vector potential and time dependent potential energy surface in the BO basis yields Eqs.~\ref{eq:expepsA}.

 \section{Mask function}
 \label{sec:mask_function}
 The mask function used in this work is defined by taking the product of two step functions, one stepping `up' on the left hand side and the other `down' on the right hand side. Defining $R_m$ to be the position of a single one of these steps, $w$ a measure of the steep-ness of the step (explain), and $s = 1$ for a step up and $s=-1$ for a step down, the \textit{single}-step function is defined by
\[ m(R,R_m,s,w) =  \begin{cases} 
     \frac{1}{2}(1-s) & R \leq R_1 \\
     \frac{1}{2} (1+s) & R\geq R_2\\
      \frac{1-s\tanh (k_1-k_2)}{2}& R_1< R< R_2       
   \end{cases}
\]
where
\begin{align}
R_{1,2} &= R_m \mp w/2 ,\notag{} \\
k_{1,2} &= \frac{w}{|R-R_{1,2}|}
\end{align}
Thus to mask away the edges of a function as done for $U_{en}$ and $\nabla \chi/\chi$ one multiplies those functions by $m(R,R_L,1,w)\times m(R,R_R,-1,w)$ where $R_L$ and $R_R$ are the left and right hand edges of the mask respectively.

\bibliography{./ref_na}

\begin{thebibliography}{39}%
\makeatletter
\providecommand \@ifxundefined [1]{%
 \@ifx{#1\undefined}
}%
\providecommand \@ifnum [1]{%
 \ifnum #1\expandafter \@firstoftwo
 \else \expandafter \@secondoftwo
 \fi
}%
\providecommand \@ifx [1]{%
 \ifx #1\expandafter \@firstoftwo
 \else \expandafter \@secondoftwo
 \fi
}%
\providecommand \natexlab [1]{#1}%
\providecommand \enquote  [1]{``#1''}%
\providecommand \bibnamefont  [1]{#1}%
\providecommand \bibfnamefont [1]{#1}%
\providecommand \citenamefont [1]{#1}%
\providecommand \href@noop [0]{\@secondoftwo}%
\providecommand \href [0]{\begingroup \@sanitize@url \@href}%
\providecommand \@href[1]{\@@startlink{#1}\@@href}%
\providecommand \@@href[1]{\endgroup#1\@@endlink}%
\providecommand \@sanitize@url [0]{\catcode `\\12\catcode `\$12\catcode
  `\&12\catcode `\#12\catcode `\^12\catcode `\_12\catcode `\%12\relax}%
\providecommand \@@startlink[1]{}%
\providecommand \@@endlink[0]{}%
\providecommand \url  [0]{\begingroup\@sanitize@url \@url }%
\providecommand \@url [1]{\endgroup\@href {#1}{\urlprefix }}%
\providecommand \urlprefix  [0]{URL }%
\providecommand \Eprint [0]{\href }%
\providecommand \doibase [0]{http://dx.doi.org/}%
\providecommand \selectlanguage [0]{\@gobble}%
\providecommand \bibinfo  [0]{\@secondoftwo}%
\providecommand \bibfield  [0]{\@secondoftwo}%
\providecommand \translation [1]{[#1]}%
\providecommand \BibitemOpen [0]{}%
\providecommand \bibitemStop [0]{}%
\providecommand \bibitemNoStop [0]{.\EOS\space}%
\providecommand \EOS [0]{\spacefactor3000\relax}%
\providecommand \BibitemShut  [1]{\csname bibitem#1\endcsname}%
\let\auto@bib@innerbib\@empty
\bibitem [{\citenamefont {Abedi}\ \emph {et~al.}(2010)\citenamefont {Abedi},
  \citenamefont {Maitra},\ and\ \citenamefont {Gross}}]{AMG10}%
  \BibitemOpen
  \bibfield  {author} {\bibinfo {author} {\bibfnamefont {A.}~\bibnamefont
  {Abedi}}, \bibinfo {author} {\bibfnamefont {N.~T.}\ \bibnamefont {Maitra}}, \
  and\ \bibinfo {author} {\bibfnamefont {E.~K.~U.}\ \bibnamefont {Gross}},\
  }\href@noop {} {\bibfield  {journal} {\bibinfo  {journal} {Phys. Rev. Lett.}\
  }\textbf {\bibinfo {volume} {105}},\ \bibinfo {pages} {123002} (\bibinfo
  {year} {2010})}\BibitemShut {NoStop}%
\bibitem [{\citenamefont {Abedi}\ \emph {et~al.}(2012)\citenamefont {Abedi},
  \citenamefont {Maitra},\ and\ \citenamefont {Gross}}]{AMG12}%
  \BibitemOpen
  \bibfield  {author} {\bibinfo {author} {\bibfnamefont {A.}~\bibnamefont
  {Abedi}}, \bibinfo {author} {\bibfnamefont {N.~T.}\ \bibnamefont {Maitra}}, \
  and\ \bibinfo {author} {\bibfnamefont {E.~K.~U.}\ \bibnamefont {Gross}},\
  }\href@noop {} {\bibfield  {journal} {\bibinfo  {journal} {J. Chem. Phys.}\
  }\textbf {\bibinfo {volume} {137}},\ \bibinfo {pages} {22A530} (\bibinfo
  {year} {2012})}\BibitemShut {NoStop}%
\bibitem [{\citenamefont {Suzuki}\ \emph {et~al.}(2014)\citenamefont {Suzuki},
  \citenamefont {Abedi}, \citenamefont {Maitra}, \citenamefont {Yamashita},\
  and\ \citenamefont {Gross}}]{SAMYG14}%
  \BibitemOpen
  \bibfield  {author} {\bibinfo {author} {\bibfnamefont {Y.}~\bibnamefont
  {Suzuki}}, \bibinfo {author} {\bibfnamefont {A.}~\bibnamefont {Abedi}},
  \bibinfo {author} {\bibfnamefont {N.~T.}\ \bibnamefont {Maitra}}, \bibinfo
  {author} {\bibfnamefont {K.}~\bibnamefont {Yamashita}}, \ and\ \bibinfo
  {author} {\bibfnamefont {E.~K.~U.}\ \bibnamefont {Gross}},\ }\href@noop {}
  {\bibfield  {journal} {\bibinfo  {journal} {Phys. Rev. A}\ }\textbf {\bibinfo
  {volume} {89}},\ \bibinfo {pages} {040501(R)} (\bibinfo {year}
  {2014})}\BibitemShut {NoStop}%
\bibitem [{\citenamefont {Khosravi}\ \emph {et~al.}(2015)\citenamefont
  {Khosravi}, \citenamefont {Abedi},\ and\ \citenamefont {Maitra}}]{KAM15}%
  \BibitemOpen
  \bibfield  {author} {\bibinfo {author} {\bibfnamefont {E.}~\bibnamefont
  {Khosravi}}, \bibinfo {author} {\bibfnamefont {A.}~\bibnamefont {Abedi}}, \
  and\ \bibinfo {author} {\bibfnamefont {N.~T.}\ \bibnamefont {Maitra}},\
  }\href@noop {} {\bibfield  {journal} {\bibinfo  {journal} {Phys. Rev. Lett.}\
  }\textbf {\bibinfo {volume} {115}},\ \bibinfo {pages} {263002} (\bibinfo
  {year} {2015})}\BibitemShut {NoStop}%
\bibitem [{\citenamefont {Abedi}\ \emph
  {et~al.}(2013{\natexlab{a}})\citenamefont {Abedi}, \citenamefont {Agostini},
  \citenamefont {Suzuki},\ and\ \citenamefont {Gross}}]{AASG13}%
  \BibitemOpen
  \bibfield  {author} {\bibinfo {author} {\bibfnamefont {A.}~\bibnamefont
  {Abedi}}, \bibinfo {author} {\bibfnamefont {F.}~\bibnamefont {Agostini}},
  \bibinfo {author} {\bibfnamefont {Y.}~\bibnamefont {Suzuki}}, \ and\ \bibinfo
  {author} {\bibfnamefont {E.~K.~U.}\ \bibnamefont {Gross}},\ }\href@noop {}
  {\bibfield  {journal} {\bibinfo  {journal} {Phys. Rev. Lett.}\ }\textbf
  {\bibinfo {volume} {110}},\ \bibinfo {pages} {263001} (\bibinfo {year}
  {2013}{\natexlab{a}})}\BibitemShut {NoStop}%
\bibitem [{\citenamefont {Min}\ \emph {et~al.}(2014)\citenamefont {Min},
  \citenamefont {Abedi}, \citenamefont {Kim},\ and\ \citenamefont
  {Gross}}]{MAKG14}%
  \BibitemOpen
  \bibfield  {author} {\bibinfo {author} {\bibfnamefont {S.~K.}\ \bibnamefont
  {Min}}, \bibinfo {author} {\bibfnamefont {A.}~\bibnamefont {Abedi}}, \bibinfo
  {author} {\bibfnamefont {K.~S.}\ \bibnamefont {Kim}}, \ and\ \bibinfo
  {author} {\bibfnamefont {E.~K.~U.}\ \bibnamefont {Gross}},\ }\href@noop {}
  {\bibfield  {journal} {\bibinfo  {journal} {Phys. Rev. Lett.}\ }\textbf
  {\bibinfo {volume} {113}},\ \bibinfo {pages} {263004} (\bibinfo {year}
  {2014})}\BibitemShut {NoStop}%
\bibitem [{\citenamefont {Curchod}\ \emph {et~al.}(2016)\citenamefont
  {Curchod}, \citenamefont {Agostini},\ and\ \citenamefont {Gross}}]{CAG16}%
  \BibitemOpen
  \bibfield  {author} {\bibinfo {author} {\bibfnamefont {B.~F.~E.}\
  \bibnamefont {Curchod}}, \bibinfo {author} {\bibfnamefont {F.}~\bibnamefont
  {Agostini}}, \ and\ \bibinfo {author} {\bibfnamefont {E.~K.~U.}\ \bibnamefont
  {Gross}},\ }\href@noop {} {\bibfield  {journal} {\bibinfo  {journal} {J.
  Chem. Phys.}\ }\textbf {\bibinfo {volume} {145}},\ \bibinfo {pages} {034103}
  (\bibinfo {year} {2016})}\BibitemShut {NoStop}%
\bibitem [{\citenamefont {Fiedlschuster}\ \emph {et~al.}(2017)\citenamefont
  {Fiedlschuster}, \citenamefont {Handt}, \citenamefont {Gross},\ and\
  \citenamefont {Schmidt}}]{FHGS17}%
  \BibitemOpen
  \bibfield  {author} {\bibinfo {author} {\bibfnamefont {T.}~\bibnamefont
  {Fiedlschuster}}, \bibinfo {author} {\bibfnamefont {J.}~\bibnamefont
  {Handt}}, \bibinfo {author} {\bibfnamefont {E.~K.~U.}\ \bibnamefont {Gross}},
  \ and\ \bibinfo {author} {\bibfnamefont {R.}~\bibnamefont {Schmidt}},\ }\href
  {\doibase 10.1103/PhysRevA.95.063424} {\bibfield  {journal} {\bibinfo
  {journal} {Phys. Rev. A}\ }\textbf {\bibinfo {volume} {95}},\ \bibinfo
  {pages} {063424} (\bibinfo {year} {2017})}\BibitemShut {NoStop}%
\bibitem [{\citenamefont {Requist}\ \emph {et~al.}(2016)\citenamefont
  {Requist}, \citenamefont {Tandetzky},\ and\ \citenamefont {Gross}}]{RTG16}%
  \BibitemOpen
  \bibfield  {author} {\bibinfo {author} {\bibfnamefont {R.}~\bibnamefont
  {Requist}}, \bibinfo {author} {\bibfnamefont {F.}~\bibnamefont {Tandetzky}},
  \ and\ \bibinfo {author} {\bibfnamefont {E.~K.~U.}\ \bibnamefont {Gross}},\
  }\href@noop {} {\bibfield  {journal} {\bibinfo  {journal} {Phys. Rev. A}\
  }\textbf {\bibinfo {volume} {93}},\ \bibinfo {pages} {042108} (\bibinfo
  {year} {2016})}\BibitemShut {NoStop}%
\bibitem [{\citenamefont {Requist}\ \emph {et~al.}(2017)\citenamefont
  {Requist}, \citenamefont {Proetto},\ and\ \citenamefont {Gross}}]{RPG17}%
  \BibitemOpen
  \bibfield  {author} {\bibinfo {author} {\bibfnamefont {R.}~\bibnamefont
  {Requist}}, \bibinfo {author} {\bibfnamefont {C.~R.}\ \bibnamefont
  {Proetto}}, \ and\ \bibinfo {author} {\bibfnamefont {E.~K.~U.}\ \bibnamefont
  {Gross}},\ }\href@noop {} {\bibfield  {journal} {\bibinfo  {journal} {Phys.
  Rev. A}\ }\textbf {\bibinfo {volume} {96}},\ \bibinfo {pages} {062503}
  (\bibinfo {year} {2017})}\BibitemShut {NoStop}%
\bibitem [{\citenamefont {Curchod}\ and\ \citenamefont
  {Agostini}(2017)}]{CA17}%
  \BibitemOpen
  \bibfield  {author} {\bibinfo {author} {\bibfnamefont {B.~F.~E.}\
  \bibnamefont {Curchod}}\ and\ \bibinfo {author} {\bibfnamefont
  {F.}~\bibnamefont {Agostini}},\ }\href@noop {} {\bibfield  {journal}
  {\bibinfo  {journal} {J. Phys. Chem. Lett.}\ }\textbf {\bibinfo {volume}
  {8}},\ \bibinfo {pages} {831} (\bibinfo {year} {2017})}\BibitemShut {NoStop}%
\bibitem [{\citenamefont {Lefebvre}(2015)}]{L15}%
  \BibitemOpen
  \bibfield  {author} {\bibinfo {author} {\bibfnamefont {R.}~\bibnamefont
  {Lefebvre}},\ }\href@noop {} {\bibfield  {journal} {\bibinfo  {journal} {J.
  Chem. Phys.}\ }\textbf {\bibinfo {volume} {142}},\ \bibinfo {pages} {074106}
  (\bibinfo {year} {2015})}\BibitemShut {NoStop}%
\bibitem [{\citenamefont {Agostini}\ and\ \citenamefont
  {Curchod}(2018)}]{Curchod_EPJB2018}%
  \BibitemOpen
  \bibfield  {author} {\bibinfo {author} {\bibfnamefont {F.}~\bibnamefont
  {Agostini}}\ and\ \bibinfo {author} {\bibfnamefont {B.~F.~E.}\ \bibnamefont
  {Curchod}},\ }\href@noop {} {\bibfield  {journal} {\bibinfo  {journal} {Euro.
  Phys. J. B}\ }\textbf {\bibinfo {volume} {91}},\ \bibinfo {pages} {141}
  (\bibinfo {year} {2018})}\BibitemShut {NoStop}%
\bibitem [{\citenamefont {Scherrer}\ \emph {et~al.}(2017)\citenamefont
  {Scherrer}, \citenamefont {Agostini}, \citenamefont {Sebastiani},
  \citenamefont {Gross},\ and\ \citenamefont {Vuilleumier}}]{SASGV17}%
  \BibitemOpen
  \bibfield  {author} {\bibinfo {author} {\bibfnamefont {A.}~\bibnamefont
  {Scherrer}}, \bibinfo {author} {\bibfnamefont {F.}~\bibnamefont {Agostini}},
  \bibinfo {author} {\bibfnamefont {D.}~\bibnamefont {Sebastiani}}, \bibinfo
  {author} {\bibfnamefont {E.~K.~U.}\ \bibnamefont {Gross}}, \ and\ \bibinfo
  {author} {\bibfnamefont {R.}~\bibnamefont {Vuilleumier}},\ }\href@noop {}
  {\bibfield  {journal} {\bibinfo  {journal} {Phys. Rev. X}\ }\textbf {\bibinfo
  {volume} {7}},\ \bibinfo {pages} {031035} (\bibinfo {year}
  {2017})}\BibitemShut {NoStop}%
\bibitem [{\citenamefont {Eich}\ and\ \citenamefont {Agostini}(2016)}]{EA16}%
  \BibitemOpen
  \bibfield  {author} {\bibinfo {author} {\bibfnamefont {F.~G.}\ \bibnamefont
  {Eich}}\ and\ \bibinfo {author} {\bibfnamefont {F.}~\bibnamefont
  {Agostini}},\ }\href@noop {} {\bibfield  {journal} {\bibinfo  {journal} {J.
  Chem. Phys.}\ }\textbf {\bibinfo {volume} {145}},\ \bibinfo {pages} {054110}
  (\bibinfo {year} {2016})}\BibitemShut {NoStop}%
\bibitem [{\citenamefont {Schild}\ \emph {et~al.}(2016)\citenamefont {Schild},
  \citenamefont {Agostini},\ and\ \citenamefont {Gross}}]{SAG16}%
  \BibitemOpen
  \bibfield  {author} {\bibinfo {author} {\bibfnamefont {A.}~\bibnamefont
  {Schild}}, \bibinfo {author} {\bibfnamefont {F.}~\bibnamefont {Agostini}}, \
  and\ \bibinfo {author} {\bibfnamefont {E.~K.~U.}\ \bibnamefont {Gross}},\
  }\href@noop {} {\bibfield  {journal} {\bibinfo  {journal} {J. Phys. Chem. A}\
  }\textbf {\bibinfo {volume} {120}},\ \bibinfo {pages} {3316} (\bibinfo {year}
  {2016})}\BibitemShut {NoStop}%
\bibitem [{\citenamefont {Requist}\ and\ \citenamefont {Gross}(2016)}]{RG16}%
  \BibitemOpen
  \bibfield  {author} {\bibinfo {author} {\bibfnamefont {R.}~\bibnamefont
  {Requist}}\ and\ \bibinfo {author} {\bibfnamefont {E.~K.~U.}\ \bibnamefont
  {Gross}},\ }\href@noop {} {\bibfield  {journal} {\bibinfo  {journal} {Phys.
  Rev. Lett.}\ }\textbf {\bibinfo {volume} {117}},\ \bibinfo {pages} {193001}
  (\bibinfo {year} {2016})}\BibitemShut {NoStop}%
\bibitem [{\citenamefont {Li}\ \emph {et~al.}(2018)\citenamefont {Li},
  \citenamefont {Requist},\ and\ \citenamefont {Gross}}]{LRG18}%
  \BibitemOpen
  \bibfield  {author} {\bibinfo {author} {\bibfnamefont {C.}~\bibnamefont
  {Li}}, \bibinfo {author} {\bibfnamefont {R.}~\bibnamefont {Requist}}, \ and\
  \bibinfo {author} {\bibfnamefont {E.~K.~U.}\ \bibnamefont {Gross}},\
  }\href@noop {} {\bibfield  {journal} {\bibinfo  {journal} {The Journal of
  Chemical Physics}\ }\textbf {\bibinfo {volume} {148}},\ \bibinfo {pages}
  {084110} (\bibinfo {year} {2018})}\BibitemShut {NoStop}%
\bibitem [{\citenamefont {Cederbaum}(2015)}]{C15}%
  \BibitemOpen
  \bibfield  {author} {\bibinfo {author} {\bibfnamefont {L.~S.}\ \bibnamefont
  {Cederbaum}},\ }\href@noop {} {\bibfield  {journal} {\bibinfo  {journal}
  {Chem. Phys.}\ }\textbf {\bibinfo {volume} {457}},\ \bibinfo {pages} {129}
  (\bibinfo {year} {2015})}\BibitemShut {NoStop}%
\bibitem [{\citenamefont {Hoffmann}\ \emph {et~al.}(2018)\citenamefont
  {Hoffmann}, \citenamefont {Appel}, \citenamefont {Rubio},\ and\ \citenamefont
  {Maitra}}]{HARM18}%
  \BibitemOpen
  \bibfield  {author} {\bibinfo {author} {\bibfnamefont {N.~M.}\ \bibnamefont
  {Hoffmann}}, \bibinfo {author} {\bibfnamefont {H.}~\bibnamefont {Appel}},
  \bibinfo {author} {\bibfnamefont {A.}~\bibnamefont {Rubio}}, \ and\ \bibinfo
  {author} {\bibfnamefont {N.~T.}\ \bibnamefont {Maitra}},\ }\href@noop {}
  {\bibfield  {journal} {\bibinfo  {journal} {The European Physical Journal B}\
  }\textbf {\bibinfo {volume} {91}},\ \bibinfo {pages} {180} (\bibinfo {year}
  {2018})}\BibitemShut {NoStop}%
\bibitem [{\citenamefont {Schild}\ and\ \citenamefont {Gross}(2017)}]{SG17}%
  \BibitemOpen
  \bibfield  {author} {\bibinfo {author} {\bibfnamefont {A.}~\bibnamefont
  {Schild}}\ and\ \bibinfo {author} {\bibfnamefont {E.~K.~U.}\ \bibnamefont
  {Gross}},\ }\href {\doibase 10.1103/PhysRevLett.118.163202} {\bibfield
  {journal} {\bibinfo  {journal} {Phys. Rev. Lett.}\ }\textbf {\bibinfo
  {volume} {118}},\ \bibinfo {pages} {163202} (\bibinfo {year}
  {2017})}\BibitemShut {NoStop}%
\bibitem [{\citenamefont {Gonze}\ \emph {et~al.}(2018)\citenamefont {Gonze},
  \citenamefont {Zhou},\ and\ \citenamefont {Reining}}]{GZR18}%
  \BibitemOpen
  \bibfield  {author} {\bibinfo {author} {\bibfnamefont {X.}~\bibnamefont
  {Gonze}}, \bibinfo {author} {\bibfnamefont {J.~S.}\ \bibnamefont {Zhou}}, \
  and\ \bibinfo {author} {\bibfnamefont {L.}~\bibnamefont {Reining}},\
  }\href@noop {} {\bibfield  {journal} {\bibinfo  {journal} {The European
  Physical Journal B}\ }\textbf {\bibinfo {volume} {91}},\ \bibinfo {pages}
  {224} (\bibinfo {year} {2018})}\BibitemShut {NoStop}%
\bibitem [{\citenamefont {Agostini}\ \emph {et~al.}(2016)\citenamefont
  {Agostini}, \citenamefont {Min}, \citenamefont {Abedi},\ and\ \citenamefont
  {Gross}}]{AMAG16}%
  \BibitemOpen
  \bibfield  {author} {\bibinfo {author} {\bibfnamefont {F.}~\bibnamefont
  {Agostini}}, \bibinfo {author} {\bibfnamefont {S.~K.}\ \bibnamefont {Min}},
  \bibinfo {author} {\bibfnamefont {A.}~\bibnamefont {Abedi}}, \ and\ \bibinfo
  {author} {\bibfnamefont {E.~K.~U.}\ \bibnamefont {Gross}},\ }\href@noop {}
  {\bibfield  {journal} {\bibinfo  {journal} {Journal of Chemical Theory and
  Computation}\ }\textbf {\bibinfo {volume} {12}},\ \bibinfo {pages} {2127}
  (\bibinfo {year} {2016})}\BibitemShut {NoStop}%
\bibitem [{\citenamefont {Min}\ \emph {et~al.}(2015)\citenamefont {Min},
  \citenamefont {Agostini},\ and\ \citenamefont {Gross}}]{MAG15}%
  \BibitemOpen
  \bibfield  {author} {\bibinfo {author} {\bibfnamefont {S.~K.}\ \bibnamefont
  {Min}}, \bibinfo {author} {\bibfnamefont {F.}~\bibnamefont {Agostini}}, \
  and\ \bibinfo {author} {\bibfnamefont {E.~K.~U.}\ \bibnamefont {Gross}},\
  }\href@noop {} {\bibfield  {journal} {\bibinfo  {journal} {Phys. Rev. Lett.}\
  }\textbf {\bibinfo {volume} {115}},\ \bibinfo {pages} {073001} (\bibinfo
  {year} {2015})}\BibitemShut {NoStop}%
\bibitem [{\citenamefont {Gossel}\ \emph {et~al.}(2018)\citenamefont {Gossel},
  \citenamefont {Agostini},\ and\ \citenamefont {Maitra}}]{GAM18}%
  \BibitemOpen
  \bibfield  {author} {\bibinfo {author} {\bibfnamefont {G.~H.}\ \bibnamefont
  {Gossel}}, \bibinfo {author} {\bibfnamefont {F.}~\bibnamefont {Agostini}}, \
  and\ \bibinfo {author} {\bibfnamefont {N.~T.}\ \bibnamefont {Maitra}},\
  }\href@noop {} {\bibfield  {journal} {\bibinfo  {journal} {Journal of
  Chemical Theory and Computation}\ }\textbf {\bibinfo {volume} {14}},\
  \bibinfo {pages} {4513} (\bibinfo {year} {2018})}\BibitemShut {NoStop}%
\bibitem [{\citenamefont {Agostini}\ \emph {et~al.}(2018)\citenamefont
  {Agostini}, \citenamefont {Tavernelli},\ and\ \citenamefont
  {Ciccotti}}]{ATC18}%
  \BibitemOpen
  \bibfield  {author} {\bibinfo {author} {\bibfnamefont {F.}~\bibnamefont
  {Agostini}}, \bibinfo {author} {\bibfnamefont {I.}~\bibnamefont
  {Tavernelli}}, \ and\ \bibinfo {author} {\bibfnamefont {G.}~\bibnamefont
  {Ciccotti}},\ }\href@noop {} {\bibfield  {journal} {\bibinfo  {journal}
  {Euro. Phys. J. B}\ }\textbf {\bibinfo {volume} {91}},\ \bibinfo {pages}
  {139} (\bibinfo {year} {2018})}\BibitemShut {NoStop}%
\bibitem [{\citenamefont {Min}\ \emph {et~al.}(2017)\citenamefont {Min},
  \citenamefont {Agostini}, \citenamefont {Tavernelli},\ and\ \citenamefont
  {Gross}}]{MATG17}%
  \BibitemOpen
  \bibfield  {author} {\bibinfo {author} {\bibfnamefont {S.~K.}\ \bibnamefont
  {Min}}, \bibinfo {author} {\bibfnamefont {F.}~\bibnamefont {Agostini}},
  \bibinfo {author} {\bibfnamefont {I.}~\bibnamefont {Tavernelli}}, \ and\
  \bibinfo {author} {\bibfnamefont {E.~K.~U.}\ \bibnamefont {Gross}},\
  }\href@noop {} {\bibfield  {journal} {\bibinfo  {journal} {The Journal of
  Physical Chemistry Letters}\ }\textbf {\bibinfo {volume} {8}},\ \bibinfo
  {pages} {3048} (\bibinfo {year} {2017})}\BibitemShut {NoStop}%
\bibitem [{\citenamefont {Gidopoulos}\ and\ \citenamefont
  {Gross}(2014)}]{GG14}%
  \BibitemOpen
  \bibfield  {author} {\bibinfo {author} {\bibfnamefont {N.~I.}\ \bibnamefont
  {Gidopoulos}}\ and\ \bibinfo {author} {\bibfnamefont {E.~K.~U.}\ \bibnamefont
  {Gross}},\ }\href@noop {} {\bibfield  {journal} {\bibinfo  {journal}
  {Philosophical Transactions of the Royal Society of London A: Mathematical,
  Physical and Engineering Sciences}\ }\textbf {\bibinfo {volume} {372}}
  (\bibinfo {year} {2014})}\BibitemShut {NoStop}%
\bibitem [{\citenamefont {Hunter}(1975)}]{H75}%
  \BibitemOpen
  \bibfield  {author} {\bibinfo {author} {\bibfnamefont {G.}~\bibnamefont
  {Hunter}},\ }\href@noop {} {\bibfield  {journal} {\bibinfo  {journal} {Int.
  J. Quantum Chem.}\ }\textbf {\bibinfo {volume} {9}},\ \bibinfo {pages} {237}
  (\bibinfo {year} {1975})}\BibitemShut {NoStop}%
\bibitem [{\citenamefont {Hunter}(1981)}]{H81}%
  \BibitemOpen
  \bibfield  {author} {\bibinfo {author} {\bibfnamefont {G.}~\bibnamefont
  {Hunter}},\ }\href@noop {} {\bibfield  {journal} {\bibinfo  {journal} {Int.
  J. Quantum Chem.}\ }\textbf {\bibinfo {volume} {19}},\ \bibinfo {pages} {755}
  (\bibinfo {year} {1981})}\BibitemShut {NoStop}%
\bibitem [{\citenamefont {Abedi}\ \emph
  {et~al.}(2013{\natexlab{b}})\citenamefont {Abedi}, \citenamefont {Maitra},\
  and\ \citenamefont {Gross}}]{AMG13}%
  \BibitemOpen
  \bibfield  {author} {\bibinfo {author} {\bibfnamefont {A.}~\bibnamefont
  {Abedi}}, \bibinfo {author} {\bibfnamefont {N.~T.}\ \bibnamefont {Maitra}}, \
  and\ \bibinfo {author} {\bibfnamefont {E.~K.~U.}\ \bibnamefont {Gross}},\
  }\href@noop {} {\bibfield  {journal} {\bibinfo  {journal} {The Journal of
  Chemical Physics}\ }\textbf {\bibinfo {volume} {139}},\ \bibinfo {pages}
  {087102} (\bibinfo {year} {2013}{\natexlab{b}})}\BibitemShut {NoStop}%
\bibitem [{\citenamefont {Alonso}\ \emph {et~al.}(2013)\citenamefont {Alonso},
  \citenamefont {Clemente-Gallardo}, \citenamefont {Echenique-Robba},\ and\
  \citenamefont {Jover-Galtier}}]{ACEJ13}%
  \BibitemOpen
  \bibfield  {author} {\bibinfo {author} {\bibfnamefont {J.~L.}\ \bibnamefont
  {Alonso}}, \bibinfo {author} {\bibfnamefont {J.}~\bibnamefont
  {Clemente-Gallardo}}, \bibinfo {author} {\bibfnamefont {P.}~\bibnamefont
  {Echenique-Robba}}, \ and\ \bibinfo {author} {\bibfnamefont {J.~A.}\
  \bibnamefont {Jover-Galtier}},\ }\href@noop {} {\bibfield  {journal}
  {\bibinfo  {journal} {The Journal of Chemical Physics}\ }\textbf {\bibinfo
  {volume} {139}},\ \bibinfo {pages} {087101} (\bibinfo {year}
  {2013})}\BibitemShut {NoStop}%
\bibitem [{\citenamefont {Agostini}\ \emph {et~al.}(2015)\citenamefont
  {Agostini}, \citenamefont {Abedi}, \citenamefont {Suzuki}, \citenamefont
  {Min}, \citenamefont {Maitra},\ and\ \citenamefont {Gross}}]{AASMMG15}%
  \BibitemOpen
  \bibfield  {author} {\bibinfo {author} {\bibfnamefont {F.}~\bibnamefont
  {Agostini}}, \bibinfo {author} {\bibfnamefont {A.}~\bibnamefont {Abedi}},
  \bibinfo {author} {\bibfnamefont {Y.}~\bibnamefont {Suzuki}}, \bibinfo
  {author} {\bibfnamefont {S.~K.}\ \bibnamefont {Min}}, \bibinfo {author}
  {\bibfnamefont {N.~T.}\ \bibnamefont {Maitra}}, \ and\ \bibinfo {author}
  {\bibfnamefont {E.~K.~U.}\ \bibnamefont {Gross}},\ }\href@noop {} {\bibfield
  {journal} {\bibinfo  {journal} {J. Chem. Phys.}\ }\textbf {\bibinfo {volume}
  {142}},\ \bibinfo {pages} {084303} (\bibinfo {year} {2015})}\BibitemShut
  {NoStop}%
\bibitem [{\citenamefont {Schwartz}(1997)}]{Schwartz97}%
  \BibitemOpen
  \bibfield  {author} {\bibinfo {author} {\bibfnamefont {C.}~\bibnamefont
  {Schwartz}},\ }\href@noop {} {\bibfield  {journal} {\bibinfo  {journal}
  {Journal of Mathematical Physics}\ }\textbf {\bibinfo {volume} {38}},\
  \bibinfo {pages} {3841} (\bibinfo {year} {1997})}\BibitemShut {NoStop}%
\bibitem [{\citenamefont {Crank}\ and\ \citenamefont {Nicolson}(1996)}]{CN96}%
  \BibitemOpen
  \bibfield  {author} {\bibinfo {author} {\bibfnamefont {J.}~\bibnamefont
  {Crank}}\ and\ \bibinfo {author} {\bibfnamefont {P.}~\bibnamefont
  {Nicolson}},\ }\href@noop {} {\bibfield  {journal} {\bibinfo  {journal}
  {Advances in Computational Mathematics}\ }\textbf {\bibinfo {volume} {6}},\
  \bibinfo {pages} {207} (\bibinfo {year} {1996})}\BibitemShut {NoStop}%
\bibitem [{\citenamefont {van~der Vorst}(1992)}]{vanderVorst}%
  \BibitemOpen
  \bibfield  {author} {\bibinfo {author} {\bibfnamefont {H.}~\bibnamefont
  {van~der Vorst}},\ }\href@noop {} {\bibfield  {journal} {\bibinfo  {journal}
  {SIAM Journal on Scientific and Statistical Computing}\ }\textbf {\bibinfo
  {volume} {13}},\ \bibinfo {pages} {631} (\bibinfo {year} {1992})}\BibitemShut
  {NoStop}%
\bibitem [{\citenamefont {Shin}\ and\ \citenamefont {Metiu}(1995)}]{SM95}%
  \BibitemOpen
  \bibfield  {author} {\bibinfo {author} {\bibfnamefont {S.}~\bibnamefont
  {Shin}}\ and\ \bibinfo {author} {\bibfnamefont {H.}~\bibnamefont {Metiu}},\
  }\href@noop {} {\bibfield  {journal} {\bibinfo  {journal} {The Journal of
  Chemical Physics}\ }\textbf {\bibinfo {volume} {102}},\ \bibinfo {pages}
  {9285} (\bibinfo {year} {1995})}\BibitemShut {NoStop}%
\bibitem [{\citenamefont {Jecko}\ \emph {et~al.}(2015)\citenamefont {Jecko},
  \citenamefont {Sutcliffe},\ and\ \citenamefont {Woolley}}]{JSW15}%
  \BibitemOpen
  \bibfield  {author} {\bibinfo {author} {\bibfnamefont {T.}~\bibnamefont
  {Jecko}}, \bibinfo {author} {\bibfnamefont {B.~T.}\ \bibnamefont
  {Sutcliffe}}, \ and\ \bibinfo {author} {\bibfnamefont {R.~G.}\ \bibnamefont
  {Woolley}},\ }\href@noop {} {\bibfield  {journal} {\bibinfo  {journal} {J.
  Phys. A: Math. Theor.}\ }\textbf {\bibinfo {volume} {48}},\ \bibinfo {pages}
  {445201} (\bibinfo {year} {2015})}\BibitemShut {NoStop}%
\bibitem [{\citenamefont {Courant}\ \emph {et~al.}(1967)\citenamefont
  {Courant}, \citenamefont {Friedrichs},\ and\ \citenamefont {Lewy}}]{CFL}%
  \BibitemOpen
  \bibfield  {author} {\bibinfo {author} {\bibfnamefont {R.}~\bibnamefont
  {Courant}}, \bibinfo {author} {\bibfnamefont {K.}~\bibnamefont {Friedrichs}},
  \ and\ \bibinfo {author} {\bibfnamefont {H.}~\bibnamefont {Lewy}},\
  }\href@noop {} {\bibfield  {journal} {\bibinfo  {journal} {IBM Journal of
  Research and Development}\ }\textbf {\bibinfo {volume} {11}},\ \bibinfo
  {pages} {215} (\bibinfo {year} {1967})}\BibitemShut {NoStop}%
\end{thebibliography}%

\end{document}